\documentclass[notitlepage,showkeys,longbibliography]{revtex4-1}
\usepackage{amsmath,amssymb}
\usepackage{graphicx}
\usepackage{dcolumn}
\usepackage{bm}

\usepackage{algorithm} 
\usepackage{algpseudocode}
\usepackage{bm}
\usepackage{bbm}
\usepackage{comment}
\usepackage[toc,page]{appendix}

\DeclareMathOperator*{\argmax}{arg\,max}
\DeclareMathOperator*{\argmin}{arg\,min}

\DeclareMathOperator{\erf}{erf}

\DeclareMathOperator{\sign}{sign}

\begin{document}


\title{Large Deviations for the Perceptron Model and Consequences for Active Learning}

\author{Hugo Cui}
 \email{hugo.cui@ens.fr}
\affiliation{ \'Ecole Normale Sup\'erieure, PSL University, Paris, France}
\author{Luca Saglietti}%
 \email{luca.saglietti@lpens.ens.fr}
\affiliation{LPENS, CNRS \& Sorbonnes Universit\'es, \'Ecole Normale Sup\'erieure, PSL University, Paris, France}

\author{Lenka Zdeborov\'a}
 \email{lenka.zdeborova@cea.fr}
\affiliation{
  Institut de Physique Th\'eorique, CNRS \& CEA \& Universit\'e Paris-Saclay, Saclay, France}

\date{\today}

\begin{abstract}%
  Active learning is a branch of machine learning that deals with problems where unlabeled data is abundant yet obtaining labels is expensive. The learning algorithm has the possibility of querying a limited number of samples to obtain the corresponding labels, subsequently used for supervised learning. In this work, we consider the task of choosing the subset of samples to be labeled from a fixed finite pool of samples. 
  We assume the pool of samples to be a random matrix and the ground truth labels to be generated by a single-layer teacher random neural network. We employ replica methods to analyze the large deviations for the accuracy achieved after supervised learning on a subset of the original pool. These large deviations then provide optimal achievable performance boundaries for any active learning algorithm. We show that the optimal learning performance can be efficiently approached by simple message-passing active learning algorithms. We also provide a comparison with the performance of some other popular active learning strategies. %
\end{abstract}

\keywords{Active learning, Large Deviations, Perceptron Model, Message-passing algorithms }

\maketitle

\section{Introduction}\label{introduction}
Supervised learning consists in presenting a parametric function (often a neural network) with a series of samples (samples) and labels, and adjusting (training) the parameters (network weights) so as to match the network output with the labels as closely as possible. Active learning (AL) is concerned with choosing the most informative samples so that the training requires the least number of labeled samples to reach the same test accuracy. Active learning is relevant in situations where the potential set of samples is large, but obtaining the labels is expensive (computationally or otherwise). There exist many strategies for active learning, see e.g. \cite{settles.tr09} for a review. In membership-based active learning \cite{membershipquery}, \cite{selectivesampling}, \cite{QueryCommittee} the algorithm is allowed to query the label of any sample, most often one it generates itself. In stream-based active learning \cite{stream} an infinite sequence of samples is presented to the learner which can decide whether or not to query its label. In pool-based active learning, which is the object of the present work, the learner can only query samples that belong to a pre-existing, fixed pool of samples. It therefore needs to choose according to some strategy which samples to query so as to have the best possible test accuracy. 

Pool-based active learning is relevant for many machine learning applications, e.g. because not every possible input vector is of relevance. A beautiful recent application of active learning is in computational chemistry \cite{interatomic} where a neural network is trained to predict inter-atomic potentials. In this case the pool of data is large and consists in all possible alloys, but not of arbitrary input vectors, and labelling is extremely expensive, as it demands resource-intensive ab-initio simulations. Consequently, only a limited number of samples can be labeled, i.e. one only possesses a certain budget for the cardinal of the training set. Another setting where a cheap large pool of input data is readily available but labelling is expensive is drug discovery \cite{Drugs}, where given a target molecule one aims to find new compounds among the pool able to bind it. Another example would be on text classification \cite{texts}, \cite{SVMText}, \cite{Text3}, where labelling a text requires non-negligible human input, while a large pool of texts is readily available on the internet.  Establishing efficient pool-based active learning procedures in this case implies to select \textit{a priori} the most informative data samples for labelling. 

Main-stream works on active learning focus on designing heuristic algorithmic strategies in a variety of settings, and analyzing the performance thereof. It is very rarely known what are the information-theoretic limitations an active learning algorithm can face and hence evaluating the distance from optimality is mostly an open question. The main contribution of the present work is to provide a toy model that is at the one hand challenging for active learning, and at the same time where the optimal performance of pool-based active learning can be computed and heuristic algorithms hence evaluated and bench-marked against the optimal solution. 
To our knowledge, this is the first work to derive optimal performance results for pool-based active learning procedures are computed. More specifically we study the random perceptron model \cite{GardnerDerrida}. The available pool of samples is assumed to be i.i.d. normal vectors, the teacher generating the labels is taken to be also a perceptron with the vector of teacher-weights having i.i.d. normal components. We compute the large deviation function for how likely one is to find a subset of the samples that leads to a given learning accuracy. Our results are based on the replica method computation of this large deviation function, that is an exact method 
(modulo the possibility of the so-called replica symmetry breaking that we are not evaluating in the present work) originating in theoretical statistical physics \cite{Replica}; \cite{Replica2}; \cite{MezardVirasoro}. Providing a rigorous proof of the obtained results or turning them into rigorous bounds would be a natural, and rather challenging, next step. In the algorithmic part of this work we benchmark several existing algorithms and also propose two new algorithms relying on the  approximate-message-passing algorithm for estimation of the label uncertainty for yet unlabeled sample, showing that they closely achieve, in the studied cases, the relevant information-theoretic limitations.

The paper is organized as follows: the problem is defined and related work discussed in section~\ref{def}. In section~\ref{mutualinfo}, we propose a measure to quantify the informativeness of given subsets of samples. In section~\ref{LD}, we derive the large deviation function over all possible subset choices and deduce performance boundaries that apply to any pool-based active learning strategy. In section~\ref{algo}, we then compare these theoretical results with the performance of existing active learning algorithms and propose two new ones, based on approximate-message-passing.

\section{Definition of the problem and related work}\label{def}

A natural modeling framework for analyzing learning processes and generalization properties is the so-called teacher-student (or planted) perceptron model \cite{review}, where the input samples are assumed to be random i.i.d. vectors, and the ground truth labels are assumed to be generated by a neural network (denoted as the teacher) belonging to the same hypothesis class as the student-neural-network. In this work we will restrict to single-layer neural networks (without hidden units) for which this setting was defined and studied in \cite{gardner1989three}. Specifically we collect the input vectors into a matrix $\bm{F} \in {\mathbb R}^{P \times N}$ where $N$ is the dimension of the input space and $P$ is the number of samples. The teacher generating the labels, called teacher perceptron, is characterized by a teacher-vector of weights ${\bf x}^0$ and produces the label vector $\bm{Y}\in\mathbb{R}^{P}$ according to ${\bf Y} = {\rm sign} ({\bf F}\cdot {\bf x}^0)$. Learning is then done using a student perceptron and consists in finding a vector $x$ so that for the training set ${\bf F}$ we have as closely as possible ${\bf Y} = {\rm sign} ({\bf F}\cdot {\bf x})$. The relevant notion of error is the test accuracy (generalization error) measuring the agreement between the teacher and the student on a new sample not presented in the training set. Since both teacher and student possess the same architecture, the training process can be rephrased in terms of an inference problem (as discussed for instance in \cite{review}): the student aims to infer the teacher weights, used to generate the labels, from the knowledge of a set of input-output associations. This scenario allows for nice geometrical insights (see for example \cite{Engels}), as the generalization properties are linked to the distance in weight space between teacher and student functions. Note that, in the case of a noiseless labelling process, the teacher-student scenario guarantees that perfect training is always possible.

Active learning was previously studied in the context of the teacher-student perceptron problem. Best known is the line of work on Query by Committee \cite{QueryCommittee,QueryCommittee2,zhou2019active}, dealing with the membership based active learning setting, i.e. where the samples are picked one by one into the training set and can be absolutely arbitrary $N$-dimensional vectors. The active learning is in that case more a strategy for designing the samples rather than one for selecting them smartly from an predefined set. In the original work \cite{QueryCommittee} the new samples are chosen so that a committee of several student-neural-networks has the maximum possible disagreement on the new sample. The paper shows that in this way one can reach a generalization error that decreases exponentially with the size of the training set, while for a random training set the generalization error can decrease only inversely proportionally to the size of the set \cite{Engels}. However, in many practical applications the possible set of samples to be picked into the training set is not arbitrarily big, e.g. not every input vector represents an encoding of a molecular structure. We hence argue that the pool-based active learning, studied in the present paper, where the samples are selected from a pre-defined set is of larger relevance to many applications. 

The theoretical part of this paper is presented for a generalization of the perceptron model, specifically the for the random teacher-student Generalized Linear Models (GLM), see e.g. \cite{Info}. An instance of a GLM is specified by a prior measure on the weights $P_{X}(\cdot)$, from which the true generative model is assumed to be sampled, and an output channel measure $P_{\mathrm{out}}(\cdot|\cdot)$, defining the generative process for the labels given the pre-activations. In the part where results of this work are presented we focus on the prototypical case of the noiseless continuous perceptron, where $P_{X}(x)=e^{-\frac{x^{2}}{2}}/\sqrt{2\pi} $ and $P_{\mathrm{out}}(y|h)=\delta(y-\mathrm{sign}(h))$ where for example $\mu$ we have $h^\mu={\bf F}^\mu \cdot {\bf x}$. Moreover, we will consider the setting where the learning model is matched to the generative model and thus the student has perfect knowledge on the correct form of the two above defined measures. 

The pool-based active learning task can now be more formally stated as follows: given a set of $N-$dimensional samples $\mathcal{S}=\{\bm{F}^{\mu}\}$ of cardinality $|\mathcal{S}|=P=\alpha N$, the goal is to select and query the labels of a subset $S\in \mathcal{S}$ of cardinality $|S|=nN$, $0<n\le\alpha$, according to some active learning criterion. We will refer to $n$ as the budget of the student. The true labels are then obtained through $y^\mu\sim P_{\mathrm{out}}(y^\mu|\bm{F}^\mu \cdot \bm{x}^0)$, \, $\bm{x}^{0}\sim P_{X}(\bm{x}^0)$. Henceforth measures with vector arguments are understood to be products over the coordinates of the corresponding scalar measures.
For technical reasons, we rely on the strong (but customary) assumption that the samples are i.i.d. Gaussian distributed, $F_{i}^{\mu}\sim\mathcal{N}(0,1)$, $\forall i\in\{1,...,N\}, ~\forall \mu \in\{1,...,P\}$. Note that, while this assumption implies that the full set $\mathcal{S}$ of input data is generally unstructured and uncorrelated, it does not prevent non-trivial correlations to appear in any smaller labeled subset $S$, selected through an active learning procedure. 

In pool-based active learning settings, it is assumed that the student has a fixed budget $n$ for building its training set, i.e. that only up to $nN$ labels can be queried for training. The active learning goal is to select, among the pool $\mathcal{S}$ of available samples, the $nN$ \textit{most informative} labels, to present to the student so that the latter achieves the best possible generalization performance.  While many criteria of informativeness have been considered in the literature, see e. g. \cite{settles.tr09}, in the teacher-student setting there exist a natural measure of informativeness, which we shall define in the next section. 

\section{The Gardner volume as a measure of optimality}\label{mutualinfo}

A natural strategy for ranking the possible subset selections is to evaluate the mutual information between the teacher vector $\bm{x^{0}}$ and each subset of labels $\bm{Y}$, conditioned on the corresponding inputs $\bm{F}$. Good selections contain larger amounts of information about the ground truth, encoded in the labels, and make the associated inference problem for the student easier. Conversely, bad selections are characterized by less informative labels. In the case of the teacher-student perceptron, where the output channel $P_{\mathrm{out}}(\cdot|\cdot)$ is completely deterministic and binary, the mutual information can be rewritten (following \cite{Info}) as 
\begin{align}
    \mathcal{I}(\bm{x}^{0};\bm{Y}|\bm{F})&=\mathcal{H}(\bm{Y}|\bm{F})-\mathcal{H}(\bm{Y}|\bm{F},\bm{x}^{0})
    =\mathcal{H}(\bm{Y}|\bm{F})  \nonumber \\
    &=-\int d\bm{Y}\int d\bm{x}^{0}\,  P_{X}(\bm{x}^{0})
    P_{\mathrm{out}}(\bm{Y}|\bm{F} \cdot \bm{x}^{0})\mathrm{ln}\int d\bm{x} P_{X}(\bm{x})P_{\mathrm{out}}(\bm{Y}|\bm{F} \cdot \bm{x}) \, . \label{mutual_to_gardner}
\end{align}
Equation (\ref{mutual_to_gardner}) allows a connection with a quantity well-known in statistical physics, the so-called Gardner volume \cite{GardnerDerrida}, \cite{Nishimori}, \cite{Engels}, denoted in the following by $v$
\begin{equation}
   \mathrm{ln} \, v \equiv \frac{1}{N} \mathbb{E}_{\bm{x^{0}},\bm{Y}}\mathrm{ln}\int d\bm{x} P_{X}(\bm{x})P_{\mathrm{out}}(\bm{Y}|\bm{F} \cdot \bm{x}) \, .
\end{equation}

The Gardner volume $v$ represents the extent of the version space \cite{Mitchell}, i.e. the entropy of hypotheses in the model class consistent with the labeled training set. This provides a natural measure of the quality of the student training. A narrower volume implies less uncertainty about the ground truth $\bm{x^{0}}$ and is thus a desirable objective in an active learning framework. We shall focus the rest of our discussion on the large deviation properties of the Gardner volume, but we invite the reader to keep in mind that this is equivalent to studying the above defined mutual information. 

There exist other natural measures of informativeness, e.g. the student generalization error $\epsilon_{g}$ and the magnetization (or teacher/student overlap) $m=\bm{x}\cdot\bm{x^{0}}/N$. In the thermodynamic limit $N\uparrow\infty$, $\epsilon_{g}$ is a decreasing function of $m$ (see the appendix~\ref{appendix:gen} for more details). Moreover we will show analytical and numerical evidence that all these measures co-vary, at least in the simple teacher-student setting studied in this work. 
A numerical check at finite $N$ of the correlation between $v$ and $m$ can also be found in appendix~\ref{appendix:numerics}.

\section{Large deviations of the Gardner volume}\label{LD}

We consider the problem of sampling labeled subsets of cardinality $n N$, $0<n\le \alpha$, from a fixed pool of data of cardinality $\alpha N$, $\alpha\sim\mathcal{O}(1)$, and study the variations in the associated Gardner volumes. We will hereby consider that, for any fixed pool and subset size, the Gardner volume probability distribution follows a large deviation principle, i.e. that there exist an exponential number $e^{N\Sigma(n,v)}$ of subsets choices that produce Gardner volumes equal to $v$. Employing a statistical physics terminology, we will refer to the rate function, $\Sigma(n,v)$, as the \textit{complexity} of labeled subsets associated to a budget $n$ and a volume $v$.

In the large $N$ limit, the overwhelming majority of subsets will thus realize a Gardner volume $v^\star$, such that $v^\star={\mathrm{argmax}_v\, }\Sigma(n,v)$. This means that, since the fluctuations around this typical value are exponentially rare, random sampling will almost certainly yield Gardner volumes extremely close to $v^\star$. However, the aim of active learning is to find strategies for accessing the atypically informative subsets (i.e., the atypically small volumes $v<v^\star$), whence the necessity of analyzing the large deviations properties of the subset selection process. 

We will here give a brief outline of the analytic computation, based on standard methods from physics of disordered systems \cite{Replica}, \cite{Replica2}, \cite{MezardVirasoro}, and refer the reader to appendices \ref{appendix:intro}, \ref{appendix:replica} and \ref{appendix:per} for a more detailed derivation. It is convenient to introduce a vector of selection variables $\{\sigma_{\mu}\}_{0\leq \mu \leq \alpha N}\in\{0,1\}^{\alpha N}$, such that $\sigma_{\mu}=1$ when the sample $\bm{F}_{\mu}\in\mathcal{S}$ is selected (and added to the labeled training set), while $\sigma_{\mu}=0$ otherwise. In this notation the selected subset $S\subset\mathcal{S}$ is easily defined as $S=\{\bm{F}_{\mu}\in\mathcal{S}|\sigma_{\mu}=1\}$. 

Since a direct computation of the complexity is not straightforward, as customary in this type of analyses \cite{PartialAnnealing} we derive it by first evaluating its Legendre transform. We introduce the (unnormalized) measure over the selection variables
\begin{equation}
    \mathbb{P}_{\beta,\phi}(\{\sigma_{\mu}\})=\left[\int d\bm{x}P_{X}(\bm{x})\prod\limits_{\mu=1}^{\alpha N}P_{\mathrm{out}}(y^{\mu}|\bm{F}^{\mu}\cdot\bm{x})^{\sigma_{\mu}}\right]^{\beta}e^{\phi \sum\limits_{\mu}\sigma_{\mu}}
\end{equation}
and the associated free entropy
\begin{equation}
    \Phi(\beta,\phi)=\mathbb{E}_{\bm{F},\bm{x^{0}}}\frac{1}{N}\mathrm{ln}\Xi=\mathbb{E}_{\bm{F},\bm{x^{0}}}\frac{1}{N}\mathrm{ln}\sum\limits_{\sigma_{\mu}}\mathbb{P}_{\beta,\phi}(\{\sigma_{\mu}\}).
\end{equation}
From a statistical physics perspective, $\Xi$ can be regarded as a grand-canonical partition function, with $\beta$ playing the role of an inverse temperature, the Gardner volume being the associated energy function, and where $\phi$ is an effective chemical potential controlling the cardinality of the selection subset, $|S|$. In the thermodynamic limit $N\uparrow\infty$, by applying the saddle-point method one can easily see that $\Phi(\beta,\phi)$ will be dominated by a subset of selection vectors $\{\sigma_{\mu}\}$ whose budget and energy, $n^\star$ and $v^\star$, are given by
\begin{equation}
\Phi(\beta,\phi)=\underset{v,n}{\mathrm{extr}}\left\{ \Sigma(n,v)+\beta\,\mathrm{ln}v +\phi\, n\right\}.
\end{equation}
Thus, inverting the Legendre transform yields the sought complexity
\begin{equation}
\Sigma(n,v)=\Phi(\beta,\phi)-\beta\,  \mathrm{ln}\, v-n\phi |_{\partial_{\beta}\Phi=\mathrm{ln}v, \partial_{\phi}\Phi=n}.
\label{eq:Legendre}
\end{equation}
At fixed budget $n$, the range of values of the volume $v$ associated to positive complexities, i.e. with $\Sigma(n,v)>0$, effectively spans all the achievable Gardner volumes for subsets of that given cardinality, agnostic of the actual strategy for selecting them. In particular, ${\mathrm{inf}}_v\{v|\Sigma(n,v)>0\}$ and  ${\mathrm{sup}}_v\{v|\Sigma(n,v)>0\}$ define the minimal and maximal Gardner volumes and provide theoretical algorithmic boundaries for all realizable active learning strategies. Note that this means that our prototypical model, albeit being idealized, constitutes a nice benchmark for comparing known pool-based active learning heuristics.

\subsection{Replica symmetric formula for the large deviations}\label{LD_RS}

In practice, the analytic evaluation of $\Phi(\beta,\phi)$ involves the computation of a quenched average of a log-partition function and is not feasible via rigorous methods. In order to perform the computation, we resort to the replica method from statistical physics \cite{Replica}, \cite{Replica2}, \cite{MezardVirasoro}, based on the identity
\begin{equation}
    \Phi(\beta,\phi)=\mathbb{E}_{\bm{F},\bm{x^{0}}}\frac{1}{N}\mathrm{ln}\, \Xi=\mathbb{E}_{\bm{F},\bm{x^{0}}}\frac{1}{N}\frac{1}{s}\underset{s\rightarrow0}{\mathrm{lim}}\Xi^{s},
\end{equation}
and the fact that for integer values $s$ the $\mathbb{E}_{\bm{F},\bm{x^{0}}}\Xi^{s}$ can be computed. 
We refer the interested reader to appendix \ref{appendix:replica}, where the computation is explicited in the more general case of a generalized linear model \cite{CS}, \cite{Info}, and then specialized to the case of interest of a teacher-student perceptron. The final analytic expression for the replica symmetric free entropy $\Phi_{\rm RS}$ in this special case reads
\begin{align}
\label{eq:Phipercept_main}
&\Phi_{\rm RS}(\beta, \phi) =\underset{\hat{m},\hat{r},\hat{q},\hat{Q},m,r,q,Q}{\mathrm{extr}}\bigg\{\frac{\beta}{2}r\hat{r}-\beta m\hat{m}-\frac{\beta(\beta-1)}{2}Q\hat{Q}+\frac{\beta^{2}}{2}q\hat{q}-\frac{\beta-1}{2}\mathrm{ln}(1+\hat{r}+\hat{Q})\nonumber\\
&~~~~~~~~~~~~~~~~~~-\frac{1}{2}\mathrm{ln}(1+\hat{r}-(\beta-1)\hat{Q}+\beta\hat{q})+\frac{\beta}{2}\frac{ \hat{q}+\hat{m}^{2}}{1+\hat{r}-(\beta-1)\hat{Q}+\beta\hat{q}}\\
&+2\alpha\int D\eta H\left(-\sqrt{\frac{m^{2}}{q-m^{2}}}\eta\right)\mathrm{ln}\left[1+e^{\phi}\int D\zeta H\left(-\frac{1}{\sqrt{r-Q}}(\sqrt{Q-q}\zeta+\sqrt{q}\eta)\right)^{\beta}\right]\bigg\},\nonumber
\end{align}
where we introduced the definitions $\int \mathcal{D}x = \int \frac{dx}{\sqrt{2\pi}} \,e^{-x^2/2}$ and $H(x) = \int_x^\infty \mathcal{D}x$.
The extremum operation entailed in the free entropy computation is performed over a set of overlap order parameters, amenable of the following geometric interpretation
\begin{itemize}
    \item $q=\frac{1}{N}\sum_i \left<\left<x_i\right>_{\boldsymbol{x}|S}\right>_S^2$, \,\,\,\,\,\,\,\, typical overlap between students with different labeled subsets.
    \item $Q=\frac{1}{N}\sum_i \left<\left<x_i\right>_{\boldsymbol{x}|S}^2 \right>_S$, \,\,\,\,\,\, typical overlap between students with the same labeled subset. 
    \item $r=\frac{1}{N}\sum_i \left<\left<x_i^2\right>_{\boldsymbol{x}|S} \right>_S$, \,\,\,\,\,\, typical norm of a student.
    \item $m=\frac{1}{N}\sum_i \left<\left<x_i x_i^0\right>_{\boldsymbol{x}|S} \right>_S$, typical magnetization of a student. 
\end{itemize}
Once the free entropy is evaluated, the complexity can be obtained via a numerical implementation of the extremization prescribed by the inverse Legendre transform (\ref{eq:Legendre}). 

We remark at this point that the presented replica calculation was obtained in the so-called replica symmetric ansatz. In general, it is possible for the replica symmetric result not to be exact, requiring replica symmetry breaking (RSB) in order to evaluate the correct free entropy $\Phi(\beta,\phi)$ \cite{MezardVirasoro}. In this model, while RSB is surely not needed close to the maximum of the complexity curves (as implied by results in \cite{Info}), it is plausibly arising for highly frustrated cases, either very large or very negative $\beta$, corresponding to the values of complexity close to zero. At the same time, the presence of RSB usually entails corrections that are very small in magnitude. We leave the (technically more challenging) investigation of the RSB solution for the large deviations for future work. 

\subsection{Large deviation results}

\begin{figure}
\begin{center}
\includegraphics[width=0.48\textwidth]{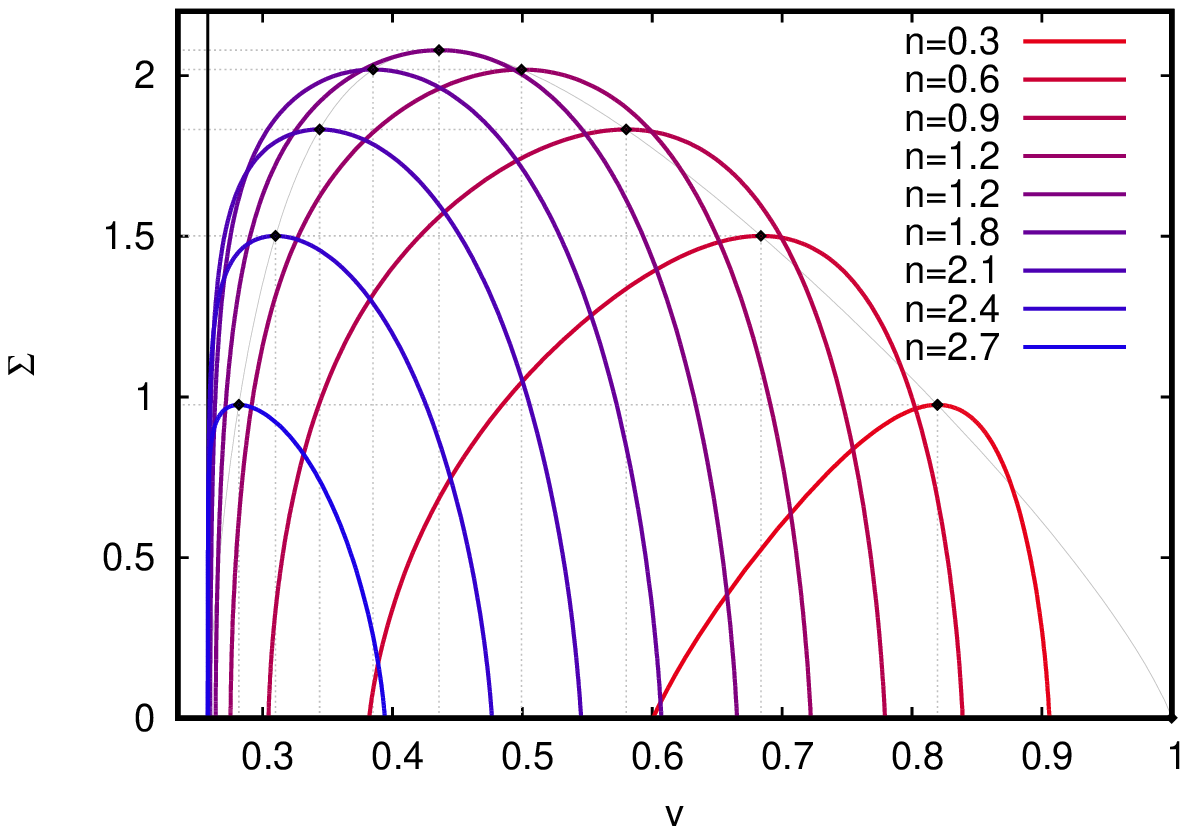} \includegraphics[width=0.48\textwidth]{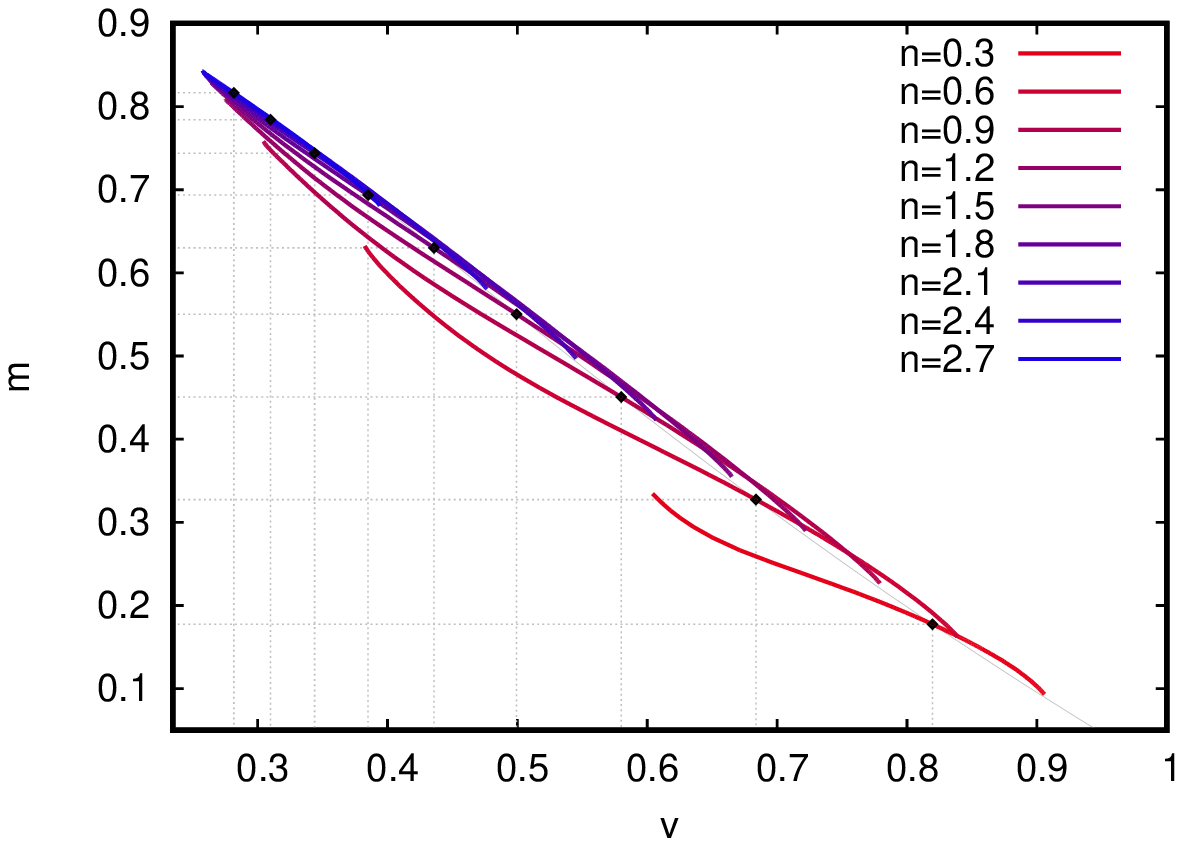}
\caption{\label{fig:energy-complexity} \label{fig:energy-magnetization}
(Left) Complexity-volume curves $\Sigma(n,v)$ for various budgets $n$, at pool size $\alpha=3$ extracted from the large deviation computations. These curves reach their maxima at a point with coordinates corresponding to the Gardner volume of randomly chosen $nN$ samples, and log-number of choices of $nN$ elements among $\alpha N$ ones.  (Right) The magnetization order parameter $m$ (in other words the teacher-student overlap) as a function of the Gardner volume $v$ for a pool of cardinality $\alpha=3$, as extracted from the large deviation computations. As is physically intuitive, smaller Gardner volumes imply larger values of the magnetization.}
\end{center}
\end{figure}

In Fig.\,\ref{fig:energy-complexity}, we show the results of the large deviation analysis at $\alpha=3$. Note that the qualitative picture is unaltered when $\alpha$ is varied (e.g., equivalent results for $\alpha=10$ are shown in appendix \ref{appendix:per}). The different curves, obtained at fixed values of the budget~$n$, show the complexity (i.e., the exponential rate of the number) of possible subset choices, $\Sigma$, that realize the corresponding Gardner volumes~$v$. As expected, the maximum of each curve is observed at $\beta=0$, and yields the typical Garner volume of a teacher-student perceptron that has learned to correctly classify $nN$ i.i.d. Gaussian input patterns. The associated complexity is simply given by the binomial distribution.  

The cases where the extremum in equation (\ref{eq:Legendre}) is realized for positive values of $\beta$ describe choices of the labeled subsets that induce atypically large Gardner volumes: these correspond to active learning scenarios where the student query is worse than random sampling. The number of possible realizations of these scenarios decreases exponentially as one approaches the right-hand extremum of where the complexity curve is positive, describing the largest possible volume at that given budget $n$. An important remark is that as soon as $\beta>0$, the statistics of the input patterns in the labeled set is no longer i.i.d., but has increasing correlations for larger $\beta$. 

On the other side, negative $\beta$ induces atypically small Gardner volumes and labeled subsets with high information content.  Again, as one spans smaller and smaller volumes the associated complexity drops, making the problem of finding these desirable subsets harder and harder. The left positive-complexity extremum of the curves in the left plot of Fig.~\ref{fig:energy-complexity} corresponds to the smallest reachable Gardner volumes. We observe in the figure that for larger values of budgets the complexity curves saturate fast very close to the smallest possible Garner volume corresponding to the Gardner volume for entire pool of samples $v(\alpha)$.

\begin{figure}
\begin{center}
\includegraphics[width=0.6\textwidth]{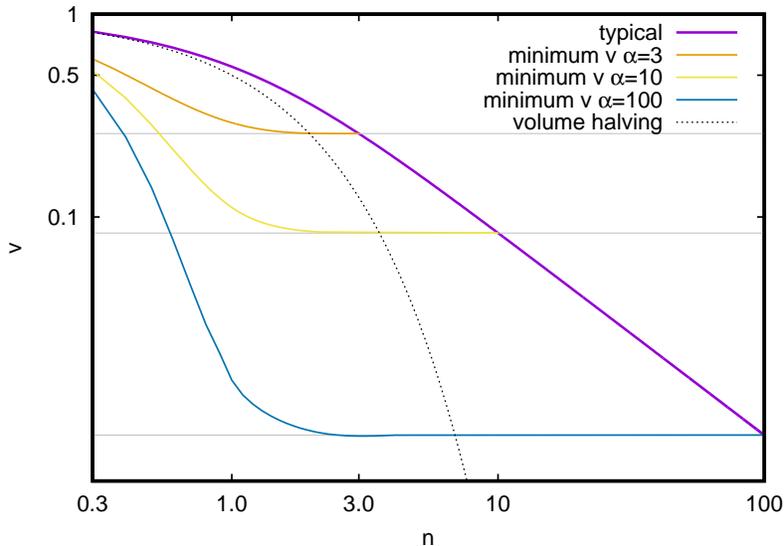}
\caption{\label{fig:pool_size_dependence}
Typical Gardner volume (purple, decreasing linearly at large $\alpha$) and information theoretically smallest achievable one (orange, yellow and blue) extracted from the large deviation computation for $\alpha\in\{3,10,100\}$. The horizontal lines depict the value of the Gardner volume corresponding to the whole pool, we see the fast saturation of the lowest volumes at these lines.  The information-theoretic volume-halving limit $2^{-n}$ for label-agnostic active learning procedures is plotted in a dotted line. We notice that the qualitative picture is essentially unchanged when $\alpha$ is varied.  
}
\end{center}
\end{figure}

In the right plot of Fig.~\ref{fig:energy-complexity}, we also show the prediction for the typical value of the magnetization, i.e. the overlap between teacher and students, as the Gardner volume is varied. As mentioned in section \ref{mutualinfo}, small Gardner volumes induce high magnetizations and thus low generalization errors.

In Fig.~\ref{fig:pool_size_dependence} the typical (purple) and corresponding minimum (orange, yellow, cyan) Gardner volumes are depicted as a function of the budget $n$ for various pool sizes $\alpha=3,~10,~100$. Note that the qualitative picture is unaltered when $\alpha$ is varied. We further observe that the minimum volume becomes very close to the Gardner volume of the entire pool of samples $v(\alpha)$ already for very small budgets $n$.


\section{Algorithmic implications}\label{algo}

\subsection{Generic considerations}

The setting investigated in this paper provides a unique chance to benchmark the algorithmic performance of any given pool-based active learning algorithm against the optimal achievable performance, and to measure how closely are the large deviations results approached. Before going to such algorithmic performance we should make a distinction between two active learning strategies 
\begin{itemize}
    \item Label-agnostic algorithms, where the student is not able to extract any knowledge on the labels. In this case, for binary labels there is a simple lower bound on the Gardner volume reachable with $nN$ samples $v\ge 2^{-n}$ which is obtained by the argument that every new sample can at best divide the current volume by a factor two \cite{QueryCommittee}. This strategy is explored in the famous query by committee active learning strategy, and the classical work \cite{QueryCommittee} argues that the volume halving can be actually achieved when unlimited set of samples is available. Plotting this volume-halving bound in Fig.~\ref{fig:pool_size_dependence} we see that even though there exit subsets leading to smaller Gardner volumes, they cannot be found in a label-agnostic way 
    \item Label-informed algorithms, where external knowledge of the labels is available and can be exploited for extracting information about which sample to choose. While in our toy setting, the additional information can only consist of disclosing the true labels, which would defy the very point of active learning, in applications with structured data the structure in the input space could be exploited (e.g., through clustering, transfer learning, etc) for making unsupervised guesses of the labels. A concrete example where external insight is available is drug discovery \cite{Drugs}, where additional information can be inferred from the presence of chemical functional groups (or absence thereof) on the molecules in the data pool. In the present work, we study whether it is possible, with full access to the labels, to devise an efficient method for finding a subset of samples that achieves close to the minimal Gardner volume (note that this is still an algorithmically non-trivial problem).  
\end{itemize}

In this section we will investigate both the label-agnostic and label-informed strategies. We will benchmark several well known active learning algorithms on the model studied in the present paper as well as design and test a new message passing based active learning algorithm. Before doing that let us describe the general strategy. 

Many of the commonly used active learning criteria rely on some form of label-uncertainty measure. Uncertainty sampling \cite{settles.tr09}, \cite{uncertainty} is an active learning scheme based on the idea of iteratively selecting and labelling data-points where the prediction of the available trained model is the least confident. In general, the computational complexity associated to this type of scheme is of order $\mathcal{O}(N^{3})$, requiring an extensive number of runs of a training algorithm (which can scale as $\mathcal{O}(N^2)$ at best). Since even training a single model per pattern addition can become expensive in the large $N$ setting, in all our numerical tests we opted for adding to the labeled set batches of $k=20$ samples instead of a single sample per iteration. We remark that, despite the $k$-fold speed-up, the observed performance deterioration is negligible. The structure of this type of algorithm is sketched in Algorithm~\ref{alg:uncertainty}.

\begin{algorithm}[H]
\caption{Uncertainty sampling} 
\label{alg:uncertainty}
\begin{algorithmic}
\State Select \emph{heuristic strategy} from Table \ref{tab:alg_tab}
\State Define batch size $k$
\State  Initialize $S\subset\mathcal{S}=\{\bm{F}_{\mu}\}_{1\le\mu\le\alpha N}$ ($|S|>0$)
\While {$|S|<nN$}
\State Obtain \emph{required estimates} given $S$
\State Obtain model predictions at data-points in $S^c$ 
\State Sort predictions according to \emph{sorting criterion} 
\State Add first $k$ elements in the sorting permutation to $S$
\EndWhile
\end{algorithmic}
\end{algorithm}

\subsection{Approximate message passing for active learning (AL-AMP)}

In general, estimating the Gardner volume on a given training set or the label-uncertainty of a new sample  is a computationally hard problem. However, in perceptrons (or more general GLMs) with i.i.d. Gaussian input data $\boldsymbol{F}$, at large system size $N$ one can rely on the estimate provided by a well known algorithm for approximate inference, Approximate Message Passing (AMP) (historically also referred to as the Thouless-Anderson-Palmer (TAP) equations, see \cite{TAP}). The AMP algorithm \cite{donoho2009message, rangan2011generalized}, \cite{review} yields (at convergence) an estimator of the posterior means, $\hat{\bm{x}}$, and variances, $\hat{\bm{\Delta}}$, thus accounting for uncertainty in the inference process including the label of a new sample. The Gardner volume~$v$ (corresponding to the so-called Bethe free entropy) can then be expressed as a simple function of the AMP fixed-point messages (see \cite{VSCS} for an example). We provide a pseudo-code of AMP in the case of the perceptron in Algorithm~\ref{alg:TAPeq}. An important remark is that when the training set is not sampled randomly from the pool, as in the active learning context, correlations can arise and AMP is no longer rigorously guaranteed to converge nor to provide a consistent estimate of the Gardner volume. In the present work, we can only argue that its employment seems to be justified \textit{a posteriori} by observing the agreement between theoretical predictions and numerical experiments for instance for the generalization error. 
\begin{algorithm}[H]
   \caption{single-instance AMP for the perceptron} 
   \label{alg:TAPeq}
\begin{algorithmic}
   \State Initialize $\bm{\hat{x}}\leftarrow 0$
   \State Initialize $\bm{\hat{\Delta}}\leftarrow 1$
   \State Initialize $\bm{g}_{\mathrm{out}}\leftarrow 0$
   \While{Convergence criterion not satisfied}
   \State $\Gamma_{\mu}^{t}\leftarrow\sum\limits_{i}(F_{i}^{\mu})^{2}\hat{\Delta}_{i}^{t-1}$
    \State $\omega^{t}_{\mu}\leftarrow\sum\limits_{i}F_{i}^{\mu}\hat{x}_{i}^{t-1}
    -\Gamma_{\mu}^{t}g_{\mathrm{out}}^{t-1}$
   \State $g_{\mathrm{out},\mu}^{t}\leftarrow\frac{y^{\mu}}{\sqrt{2\pi \Gamma_{\mu}^{t}}}\frac{e^{-\frac{(\omega_{\mu}^{t})^{2}}{2\Gamma_{\mu}^{t}}}}{H\left(-\frac{y^{\mu}\omega_{\mu}^{t}}{\sqrt{\Gamma^{t}_{\mu}}}\right)}$
    \State $(\Sigma_{i}^{t})^{-1}\leftarrow-\sum\limits_{\mu}(F_{i}^{\mu})^{2}
    ( -\frac{\omega^{t}}{V^{t}}g_{\mathrm{out}}^{t}
    -(g_{\mathrm{out}}^{t})^{2})$
    \State$ R_{i}^{t}\leftarrow\hat{x}_{i}^{t-1}+\Sigma_{i}^{t}
    \sum\limits_{\mu}F_{i}^{\mu}
    g_{\mathrm{out}}^{t}$
    \State $\hat{x}_{i}^{t}\leftarrow\frac{R_{i}^{t}}{1+\Sigma^{t}_{i}}$
    \State $\hat{\Delta}_{i}^{t}\leftarrow\frac{\Sigma_{i}^{t}}{1+\Sigma^{t}_{i}}$
   
	   \EndWhile
\end{algorithmic}
\end{algorithm}

We use the AMP algorithm to introduce a new uncertainty sampling procedure relying on the information contained in the AMP messages, denoted as AL-AMP in the following. At each iteration, the single-instance AMP equations are run on the current training subset to yield posterior mean estimate $\hat{\bm{x}}$ and variance $\bm{\hat{\Delta}}$. These quantities can then be used to evaluate, for all the unlabeled samples, the output magnetization (i.e. the Bayesian prediction) defined as
\begin{equation}
m_{\mathrm{out}}^\mu=\erf(\omega^\mu/\sqrt{2\Gamma^\mu})~~~~~~~\forall\mu|\sigma_\mu=0, \label{new_labels}
\end{equation} {}
where we introduced the output overlaps $\bm{\omega}=\bm{F}'\cdot\hat{\bm{x}}$ and variances $\bm{\Gamma}=(\bm{F}'\odot \bm{F}')\cdot\hat{\bm{\Delta}}$, where $\odot$ is the component-wise product. The output magnetizations correspond to the weighted output average over all the estimators contained in the current version space, and their magnitude represents the overall confidence in the classification of the still unlabeled samples. This means that AMP provides an extremely efficient way of obtaining the information on uncertainty. 
The specifics of the algorithm can be found in Tab.~\ref{tab:alg_tab}. 

We also explore numerically the label-informed active learning regime introduced in the previous section.  We consider its limiting case by introducing the informed AL-AMP strategy, which can fully access the true labels $\bm{Y}$ in order to query the samples $\bm{F}^{\mu}$ whose output magnetisation $m_{\mathrm{out}}^\mu$ (\ref{new_labels}) is maximally distant from the correct classification $y^{\mu}$. This selection process can iteratively reduce the Gardner volume of factors larger than $2$. Again, the relevant specifics of informed AL-AMP algorithm are detailed in Tab.\,\ref{tab:alg_tab}.

\renewcommand{\arraystretch}{1.5}
\begin{center}
\begin{table}
\begin{tabular}{ |p{4cm}||p{3cm}|p{6cm}|  } 
 \hline
 \multicolumn{3}{|c|}{Uncertainty sampling strategies} \\  [0.5ex] 
 \hline
 Heuristic &Required estimates &Sorting criterion\\ [0.5ex]
 \hline 
 \emph{\rm Agnostic AL-AMP}     & $\hat{\bm{x}}_\text{AMP}, \hat{\bm{\Delta}}_\text{AMP}$ & $\argmin_\mu \left|\erf\left(\frac{\bm{F}'^\mu\hat{\bm{x}}_\text{AMP}}{\sqrt{2(\bm{F}'^\mu)^2\hat{\bm{\Delta}}_\text{AMP}}}\right)\right|$\\ 
 \emph{\rm Informed AL-AMP} &  $\hat{\bm{x}}_\text{AMP}, \hat{\bm{\Delta}}_\text{AMP}$ & $ \argmax_\mu \left|y^\mu - \erf\left(\frac{\bm{F}'^\mu\hat{\bm{x}}_\text{AMP}}{\sqrt{2(\bm{F}'^\mu)^2\hat{\bm{\Delta}}_\text{AMP}}}\right)\right|$\\
 \emph{\rm Query by committee}   & $\{\bm{x}_\text{SGD}^k\}_{k=1}^{K}$ &  $\argmin_\mu |\sum_{k=1}^{K} \sign\left(\bm{F}'^\mu\cdot \bm{x}_\text{SGD}^k\right)|$\\
 \emph{\rm Logistic regression} & $\bm{x_\text{log}}$ &  $\argmin_\mu \left|\bm{F}'^\mu\cdot \bm{x}_\text{log}\right|$  \\ 
 \emph{\rm Perceptron learning} & $\bm{x_\text{perc}}$ &  $\argmin_\mu \left|\bm{F}'^\mu\cdot \bm{x}_\text{perc}\right|$  \\
 \hline
\end{tabular}
\caption{\label{tab:alg_tab} Table summarizing the specifics of the uncertainty sampling strategies considered in this paper.}
\end{table}
\end{center}

\subsection{Other tested measures of uncertainty}

One of the widely used uncertainty sampling procedure is the so-called Query by Committee (QBC) strategy \cite{QueryCommittee}, \cite{QueryCommittee2}. In QBC, at each time step, a committee of $K$ students is to be sampled from the version space (e.g., via the Gibbs algorithm). The committee is then employed to choose the labels to be queried, by identifying the samples where maximum disagreement in the committee members outputs is observed. The QBC algorithm was introduced as a proxy for doing bisection, i.e. cutting version space into two equal-volume halves. As already mentioned, this constitutes the optimal information gain in an label-agnostic setting \cite{greedy}. Note that, however, the QBC procedure can achieve volume-halving only in the infinite-size committee limit, $K\uparrow\infty$, with uniform version space sampling and with availability of infinitely many samples. Obviously, running a large number $K$ of ergodic Gibbs sampling procedures quickly becomes computationally unfeasible. Moreover in the pool-based active learning the pool of samples is limited. In order to allow comparison with other strategies at finite sizes, we approximated the uniform sampling with a set of greedy optimization procedures (e.g., stochastic gradient descent) from random initialization conditions, checking numerically that this yields a committee of students reasonably spread out in version space. It is possible to ensure a greater coverage of the version space by performing a short Monte-Carlo random walk for each committee member. The effect has been found to be small for computationally reasonable lengths of walk.

We also implemented an alternative uncertainty sampling strategy, relying on a single training procedure (e.g., training with the perceptron algorithm or logistic regression) per iteration: in this case, the uncertainty information is extracted from the magnitude of the pre-activations measured at the unlabeled samples after each training cycle. This strategy implements the intuitive geometric idea of looking for the samples that are most orthogonal to the available updated estimator, which are more likely to halve the version space independently of the value of the true label.

\subsection{Algorithmic results}

\begin{figure}[ht]
\begin{center}
\includegraphics[width=0.48\textwidth]{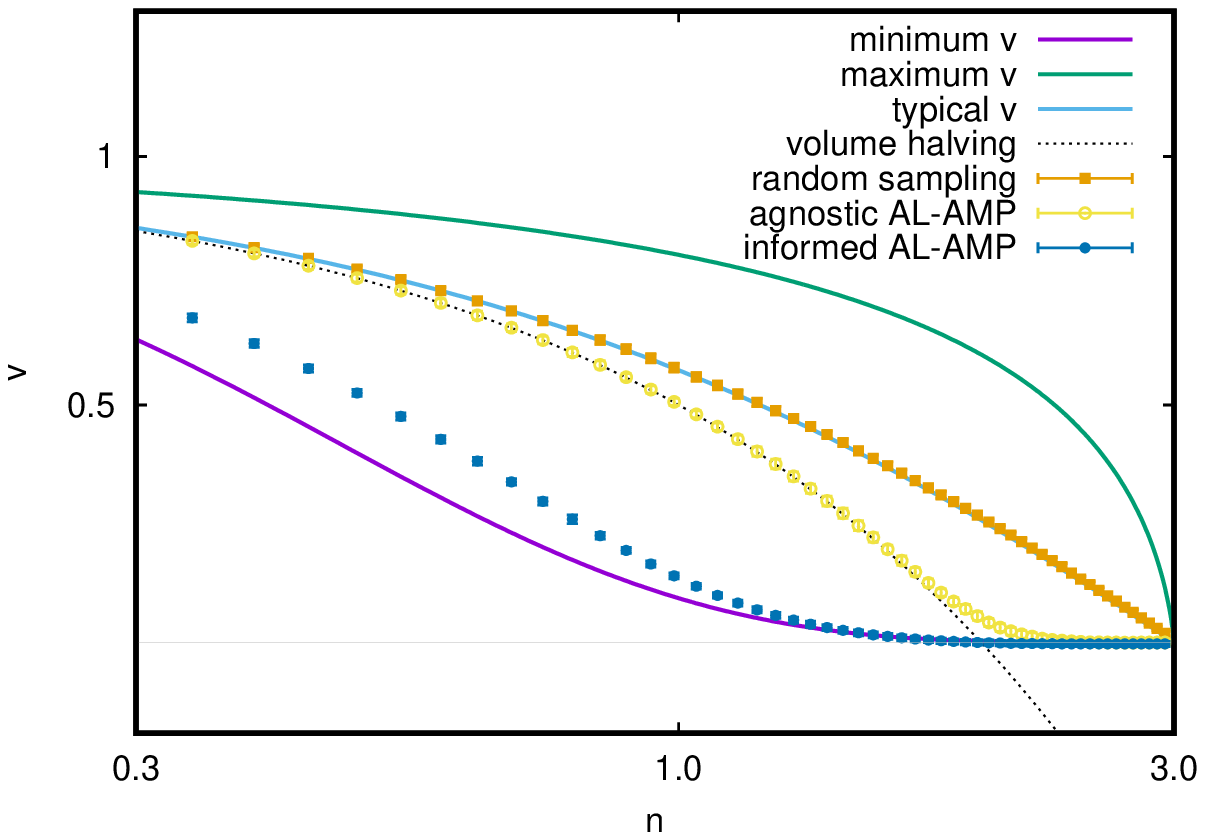}
\includegraphics[width=0.48\textwidth]{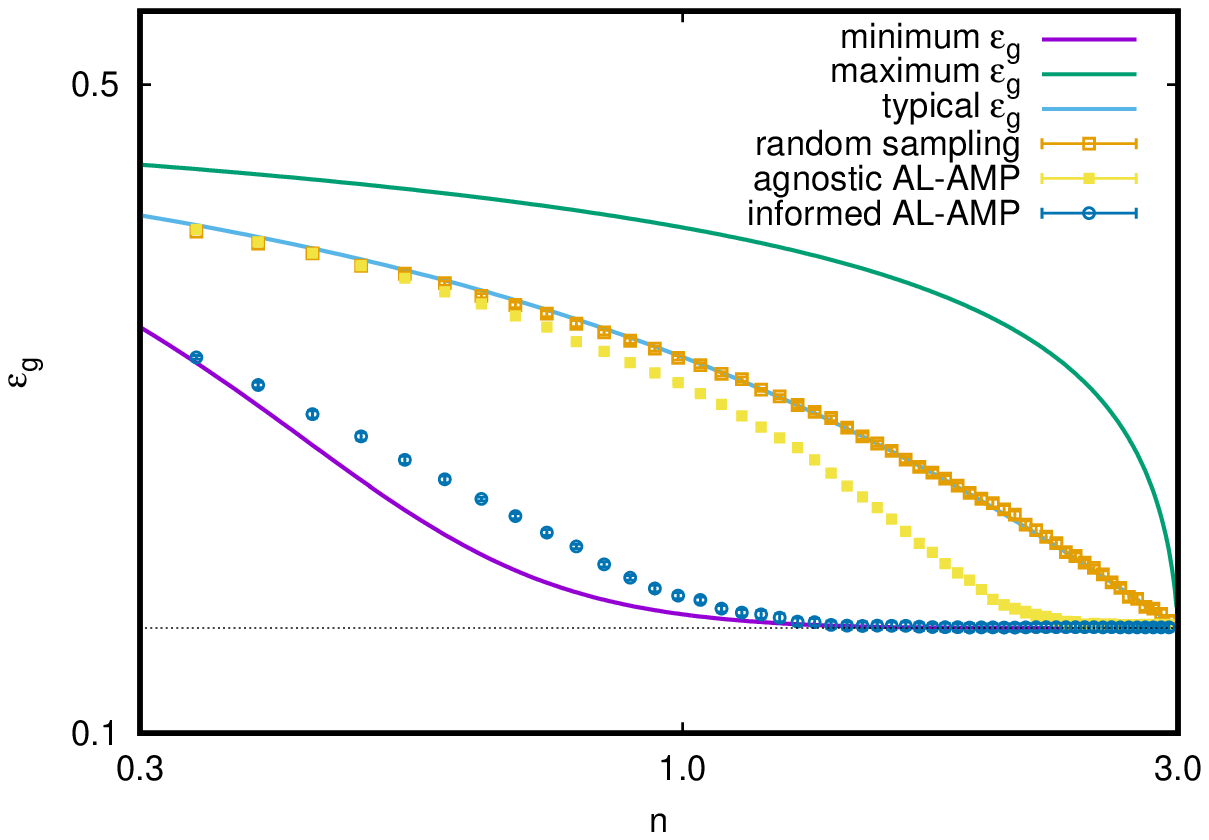}
\caption{\label{fig:n_vs_v_bounds}
(Left) Performance of the label-agnostic (yellow circles) and label-informed (blue circles) AL-AMP, plotted together with the minimum and maximum values of the Gardner volume extracted from the large deviation computation (purple and green) and the volume-halving curve (dotted black). For comparison we also plot the typical Gardner volume (cyan) and the one obtained by random sampling (orange squares). Numerical experiments were run for system size $N=2\cdot10^3$ and pool size $\alpha=3$. For each algorithmic performance curve the average over $10$ samples is presented. Fluctuations were found to be negligible around the average and are not shown. (Right) The same plot with the Gardner volume replaced by the Bayesian test accuracy, derived in Appendix \ref{appendix:gen}. For the AL-AMP algorithm the accuracy is evaluated using a test set of size $P_{\mathrm{test}}=5\cdot10^4$. The qualitative picture is very similar to the one for the Gardner volume curves (left), once more confirming that Gardner volumes and generalization errors both constitute good  measures for informativeness.
}
\end{center}
\end{figure}

\begin{figure}
\begin{center}
\includegraphics[width=0.48\textwidth]{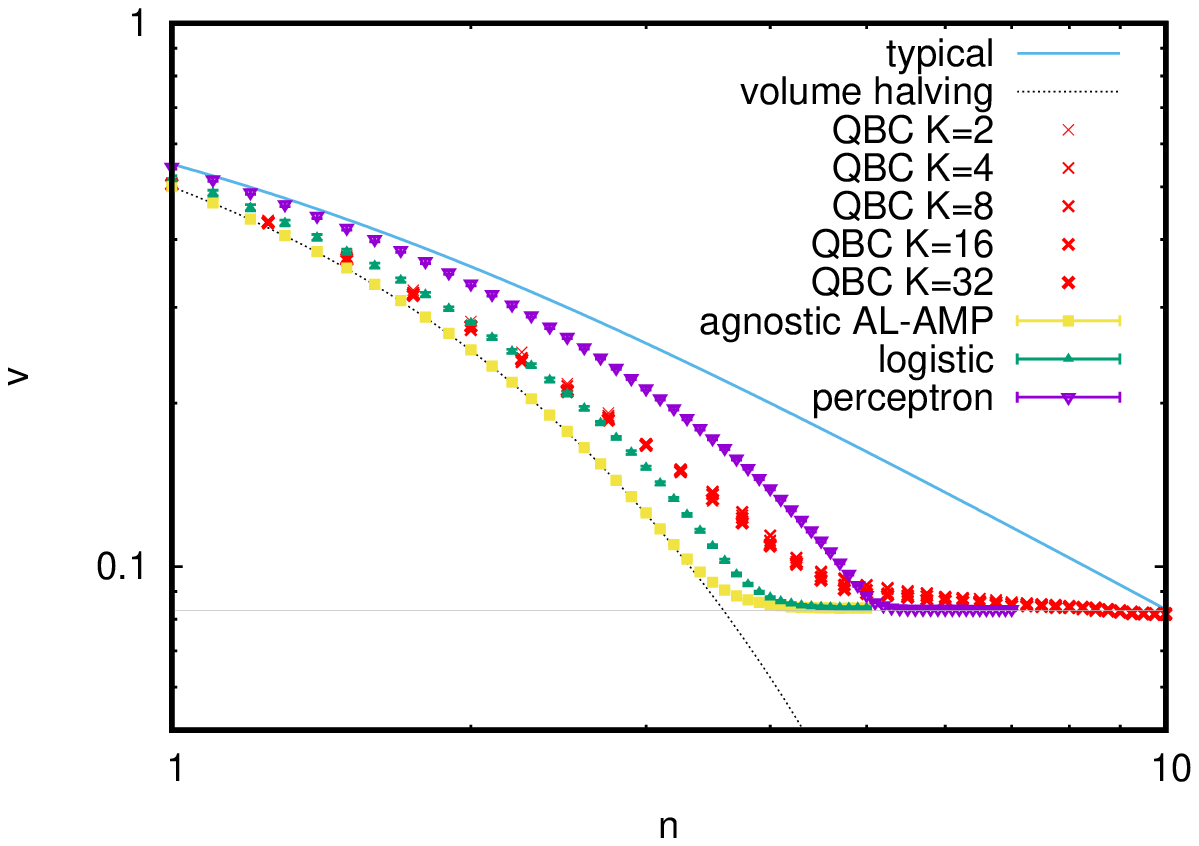}
\includegraphics[width=0.48\textwidth]{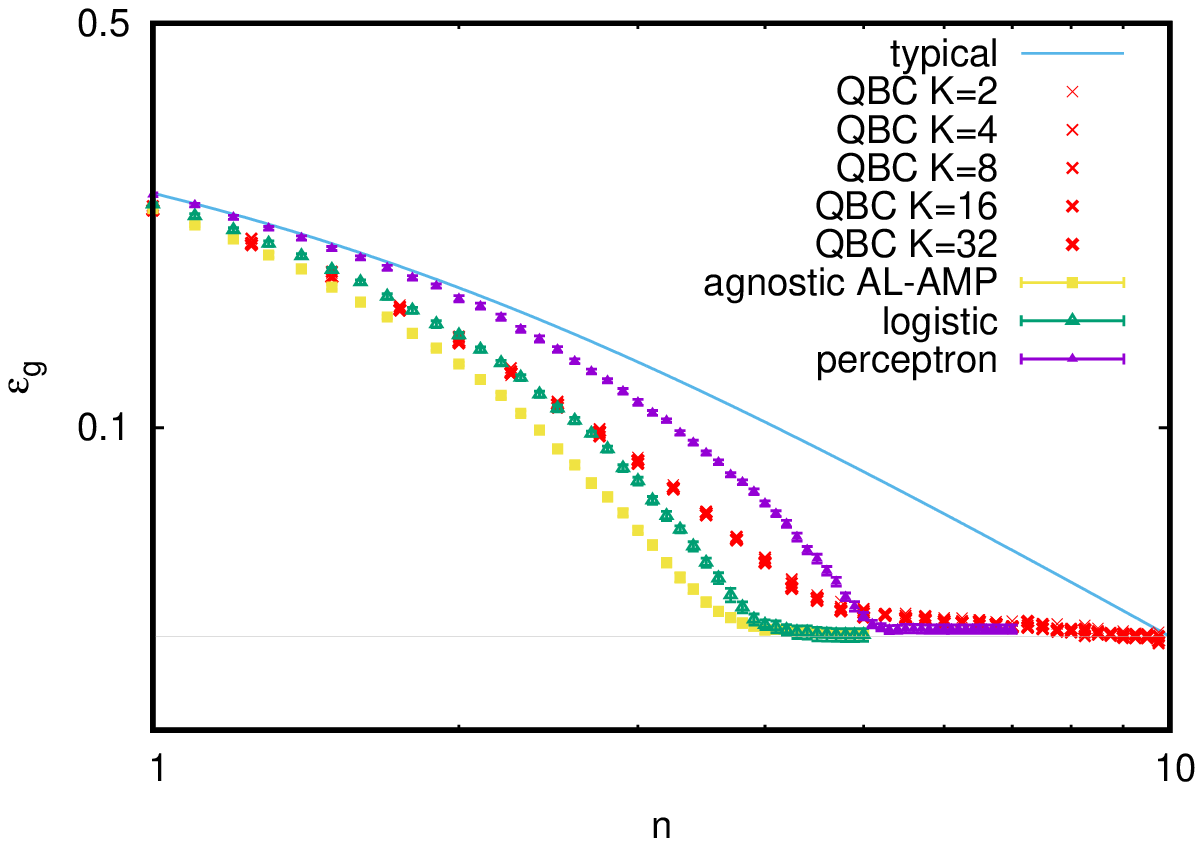}
\caption{\label{fig:other_heuristics}
(Left) Performance of the label-agnostic algorithms presented in Tab.~\ref{tab:alg_tab} is plotted against the budget $n$ and compared to the volume-halving lower bound. Experiments were performed at system size $N=2\cdot10^3$, and pool size $\alpha=10$. For each algorithm the average over $10$ samples is presented. Note that error bars are smaller than marker size. (Left) (Bayesian) test accuracy  of the same heuristics for various budgets $n$. The test set size was chosen to be $P_{\mathrm{test}}=5\cdot10^4$. In blue the Bayesian test accuracy for a typical subset, see appendix \ref{appendix:gen}. Again, the qualitative picture is unchanged going from the Gardner volume to the test accuracy.
}
\end{center}
\end{figure}

In Fig.~\ref{fig:n_vs_v_bounds}, we compare the minimum Gardner volume obtained from the large deviation calculation with the algorithmic performance obtained on synthetic data at finite size, $N=2\cdot10^3$, by the AL-AMP algorithms detailed in Algorithm \ref{alg:uncertainty} and Tab. \ref{tab:alg_tab}. The data-pool size is fixed to $\alpha=3$. The large deviation analysis yields values for the minimum and maximum achievable Gardner volumes at any budget $n$. We compare the algorithmic results also with the prediction for the typical case and with the volume-halving curve $2^{-n}$. Since in the considered pool-based setting the volume-halving performance cannot be achieved for volumes smaller than the Gardner volume corresponding to the entire pool $v(\alpha)$, the relevant volume-halving bound should be more precisely ${\rm max}(2^{-n},v(\alpha))$. Random sampling displays good agreement with the expected typical volumes. Most notably, the label-agnostic AL-AMP algorithm tightly follows the volume-halving bound ${\rm max}(2^{-n},v(\alpha))$, thus reaching close to optimal possible performance. 
Since for large $\alpha$ the behaviour of $v(\alpha) ={\rm const.} / \alpha$ \cite{Engels} we conclude that the AL-AMP algorithm will reach close to minimum possible Gardner volumes for a budget $n \sim {\cal O}[\log(\alpha)]$. We thus obtain an exponential reduction in the number of samples even in the pool-based active learning similarly to the original Query by Committee work \cite{QueryCommittee}.  


The label-informed AL-AMP also approaches the theoretically minimal volume but not as closely. 
We remark that an important limit of the AL-AMP algorithm comes from the fact that AMP is not guaranteed to provide good estimators (or converge at all) with correlated data. For example, in the numerical experiments for obtaining the informed AL-AMP curve, we had to resort to mild damping schemes in the message-passing to allow fixed-points being reached. This effect was stronger for the label-informed algorithm than for the label-agnostic one. 

In Fig.\,\ref{fig:other_heuristics}, we provide a numerical comparison of the performance of the agnostic AL-AMP and the other above mentioned label-agnostic active learning algorithms. The finite size experiments were run at $N=2.10^3$, while here we set $\alpha=10$. Note that, while the mentioned different active learning strategies where employed for selecting the labeled subset, in all cases supervised learning and the related performance estimates were obtained by running AMP.
In the plot, we can see that, while AL-AMP is able to extract very close to the maximum amount of information from each query (one bit per pattern, until the volume $v(\alpha)$ is saturated), other heuristics with the same computational complexity are sub-optimal. In particular, in the simplified query by committee procedure we observe that increasing the size $K$ of the committee does not yield very noticeable change in its performance, most probably because the committee cannot cover a sufficient portion of the version space if the computational cost is to be kept reasonable. On the other hand, using the information of the magnitude pre-activations allows better performance while being also more time-efficient, since only a single perceptron, rather than a committee thereof, has to be trained at each step. The logistic loss allows a rather good performance, close to that of AL-AMP, while the uncertainty sampling with the perceptron algorithm yields a mitigated performance.   

We leave a more systematic bench-marking of the many existing strategies for future work, stressing the fact that, while there certainly exist more involved procedures that can yield better performance than the  presented heuristics, the absolute performance bounds still apply, agnostic of the implemented active learning strategy.

\section{Conclusions}\label{Conclusion}
Using the replica method for large deviation calculation of the Gardner volume, we computed for the teacher-student perceptron model the minimum Gardner volume (equivalently, maximum mutual information) achievable by selecting a subset with fixed cardinality from a pre-existing pool of i.i.d. normal samples. We evaluated the large deviation function based on the replica symmetric assumption; checking for replica symmetry breaking and evaluating the eventual corrections to the presented results is left for future work, as well as rigorous establishment of the presented results. 
Our result for the information-theoretic limit of pool-based active learning in this setting  complements the already known volume-halving bound for label-agnostic strategies. 
We hope our result  may serve as a guideline to benchmark future heuristic algorithms on the present model, while our modus operandi regarding the derivation of the large deviations may help for future endeavour in theoretical analysis of active learning in more realistic settings. We presented the performance of some known heuristics, plus we suggested the AL-AMP algorithms to perform the uncertainty based active learning. We show numerically that on the present model the label-agnostic AL-AMP algorithm performs very close to the optimal bound, thus being able to achieve accuracy corresponding to the entire pool of samples with exponentially fewer samples.


\begin{acknowledgements}
We want to thank Guilhem Semerjian for clarifying discussions in early stages on this work. This work is supported by the ERC under the European Union’s Horizon 2020 Research and Innovation Program
714608-SMiLe.
\end{acknowledgements}

\bibliography{bibliography}
\newpage

\appendix
\section{Notations and Large-Deviation (LD) measure for GLMs}
\label{appendix:intro}
In this appendix we set the replica calculation in the more general setting of a generalized linear model (GLM) with arbitrary teacher/student prior/posterior. We allow the inference to be mismatched, i.e. we allow the teacher and student measures to be different. The specialization to the particular case of the teacher-student perceptron with no mismatch (Bayes-optimal) will be carried out in appendix \ref{appendix:per}. Our computation borrows from \cite{CS} and \cite{Info} which study the simple measure. Because we study large deviations, our formalism has some semblance with one-step Replica Breaking (1RSB) equations, a discussion of which can be found for example in \cite{ASP}, \cite{RSB}, \cite{MezardVirasoro}.
\subsection{Definition of the problem}

We consider a student GLM (\cite{CS}) with $N$-dimensional weights learning from $\alpha N$ samples $\bm{F}^{\mu}\in\mathbb{R}^{N}$ stacked in a matrix $\bm{F}\in\mathcal{M}_{\alpha N,N}(\mathbb{R})$ and the corresponding labels $y^{\mu}$ stacked into $\bm{Y}\in \mathbb{R}^{\alpha N}$. We assume the student-teacher (or planted) setting (\cite{review}), where the labels are generated by the ground truth (teacher weights) $\bm{x}^{0}$ with the channel measure $\overline{P_{\mathrm{out}}}(y^{\mu}|\bm{F}^{\mu}\cdot\bm{x^{0}})$. The teacher weight itself is drawn with prior $\overline{P_{X}}(\cdot)$. Given $\bm{F}$ and $\bm{Y}$, the student perceptron is trained so that its own weight vector $\bm{x}$ tries to match the ground truth $\bm{x}^{0}$. The inference is carried out with the student prior $P_{X}(\cdot)$ and posterior $P_{\mathrm{out}}(\bm{Y}|\bm{F}\cdot\bm{x})$. Note that the cases where $P_{X}(\cdot)\ne\overline{P_{X}}(\cdot)$ or $P_{\mathrm{out}}(\cdot)\ne\overline{P_{\mathrm{out}}}(\cdot)$ mean that the student ignores the precise Markov chain wherefrom the labels are generated, as discussed in section \ref{def}, see also \cite{review}. The likelihood that a vector $\bm{x}$ is the ground truth vector is then $P_{X}(\bm{x})P_{\mathrm{out}}(\bm{Y}|\bm{F}\cdot\bm{x})$. A reasonable measure of the average lack of accuracy of the student's guess is then given by the Gardner volume, viz. the partition function associated with the likelihood
\begin{equation}
v=\left(\int d\bm{x}P_{X}(\bm{x})P_{\mathrm{out}}(\bm{Y}|\bm{F}\cdot\bm{x})\right)^{\frac{1}{N}} .
\end{equation}
The smaller $v$, the easier the student inference, see section \ref{def} in the main text. The validity of the Gardner volume as a measure of informativeness is justified for the Bayes-optimal perceptron in section \ref{mutualinfo} of the main text.

We consider here pool-based active learning, where only a subset $S$ of the pool $\mathcal{S}=\{\bm{F}^{\mu}\}_{1\le \mu\le\alpha N}$ is used for training. The choice of subset can be conveniently parametrized by the Boolean $\sigma_{\mu}\in\{0,1\}$, where $\sigma_{\mu}=1$ means sample $\bm{F}_{\mu}$ is used, while $\sigma_{\mu}=0$ means  $\bm{F}_{\mu}$ is not selected. For a given budget $0\leq n =\frac{1}{N}|S|=\frac{1}{N}\sum\limits_{\mu=1}^{\alpha N}\sigma_{\mu}\leq \alpha$, we intend to find the selection $S$ that minimizes the Gardner volume $v(\sigma_{\mu})$, viz. that allows the best student guess. To do this we shall compute the complexity $\Sigma(n,v)$, with $e^{N\Sigma(n,v)}$ the number of ways to select $nN$ samples so that the Gardner volume associated with the training of the student is $v$, as in section \ref{LD} of the main text.
\subsection{Assumption}
To simplify, the samples are taken to be identically and independently distributed according to a normal distribution $\forall(i,\mu),~F^{\mu}_{i}\overset{d}{=}\mathcal{N}(0,\frac{1}{\sqrt{N}})$. Moreover all measures over vectors are assumed to be separable, that is factorizable as a product of identical measures over the components. Notation-wise  $P_{X}(\bm{x})$ for example is therefore understood to mean $\prod\limits_{i=1}^{N}P_{X}(x_{i})$.
\subsection{LD measure}
The goal is to compute the averaged log partition function (free entropy in statistical physics terms)
\begin{equation}
\label{eq:GLMmeasure}
\Phi(\beta,\phi)=\mathbb{E}_{\bm{F},\bm{Y},\bm{x^{0}}}\frac{1}{N}\mathrm{ln}\Xi=\mathbb{E}_{\bm{F},\bm{Y},\bm{x^{0}}}\frac{1}{N}\mathrm{ln}\sum\limits_{\sigma_{\mu}}\left[\int d\bm{x}P_{X}(\bm{x})\prod\limits_{\mu=1}^{\alpha N}P_{\mathrm{out}}(y^{\mu}|\bm{F}^{\mu}\cdot\bm{x})^{\sigma_{\mu}}\right]^{\beta}e^{\phi \sum\limits_{\mu}\sigma_{\mu}}.\end{equation}
$\beta$ can be seen as an inverse temperature and $\phi$ as a chemical potential, see section \ref{LD} in the main text. The reason why we compute $\Phi(\beta,\phi)$ is that this quantity is the Legendre transform of the complexity $\Sigma(n,v)$
\begin{equation}
\Phi(\beta,\phi)=\underset{v,n}{\mathrm{extr}}\left\{ \Sigma(n,v)+\beta \mathrm{ln}v +\phi n\right\}.
\end{equation}
Inverting the Legendre transform is then straightforward and yields $\Sigma(n,v)$
\begin{equation}
\Sigma(n,v)=\Phi(\beta,\phi)-\beta \mathrm{ln}v-n\phi |_{\partial_{\beta}\Phi=\mathrm{ln}v, \partial_{\phi}\Phi=n}.
\end{equation}
The spectrum of values of $v$ such that $\Sigma(n,v)>0$ for any given fixed $n$ corresponds to all achievable Gardner volumes for a budget $n$. In particular, $\underset{v}{\mathrm{inf}}\{v|\Sigma(n,v)>0\}$ is the minimal Gardner volume when the $nN$ samples are chosen in an optimal way. Contrariwise, $\underset{v}{\mathrm{sup}}\{v|\Sigma(n,v)>0\}$ is the maximal Gardner volume when the $nN$ samples are chosen in the least informative way for the student,  so that the student inference problem is hardest. Finally note that the selection variables $\{\sigma_{\mu}\}$ play in the grand-canonical partition function (\ref{eq:GLMmeasure}) the role of an annealed disorder in disordered systems terminology \cite{MezardVirasoro}, and shall be sometimes referred to as such in the following. 

\section{Replica computation}
\label{appendix:replica}
\subsection{Replica trick}
The standard way of taking care of the logarithm in equation (\ref{eq:GLMmeasure}) is the replica trick (\cite{Replica}, \cite{Replica2}, \cite{MezardVirasoro}),
\begin{equation}
\label{eq:replica}
\Phi(\beta,\phi)=\underset{s\rightarrow0}{\mathrm{lim}}\frac{1}{s}\mathbb{E}_{\bm{F},\bm{Y},\bm{x^{0}}}\Xi^{s}.
\end{equation}
To compute $\mathbb{E}_{\bm{F},\bm{Y},\bm{x^{0}}}\Xi^{s}$, one needs to further replicate $\beta$ times to care for the power $\beta$ involved in the summand in equation (\ref{eq:GLMmeasure})
\begin{align}
\mathbb{E}_{\bm{F},\bm{x^{0}}}\Xi^{s}&=\mathbb{E}_{\bm{F}}\int d\bm{x}^{0}\overline{P_{X}}(\bm{x}^{0})\int d\bm{y}\prod\limits_{\mu=1}^{\alpha N}\overline{P_{\mathrm{out}}}(y^{\mu}|\bm{F}^{\mu}\cdot\bm{x}^{0})\nonumber\\&\sum\limits_{S_{a}^{\mu}}\int  \prod\limits_{a=1}^{s}\prod\limits_{\alpha=1}^{\beta}(d\bm{x}^{a\alpha}P_{X}(\bm{x}^{a\alpha}))\prod\limits_{a\alpha}\prod\limits_{\mu}P_{\mathrm{out}}(y^{\mu}|\bm{F}^{\mu}\cdot\bm{x}^{a\alpha})^{\sigma^{a}_{\mu}}e^{\phi \sigma^{a}_{\mu}}\nonumber\\
&=\mathbb{E}_{\bm{F}}\sum\limits_{S_{a}^{\mu}}\int \prod\limits_{a\alpha}(d\bm{x}^{a\alpha}P_{X}(\bm{x}^{a\alpha}))e^{-\phi}\prod\limits_{a\alpha}\prod\limits_{\mu}P_{\mathrm{out}}(y^{\mu}|\bm{F}^{\mu}\cdot\bm{x}^{a\alpha})^{\sigma^{a}_{\mu}}e^{\phi \sigma^{a}_{\mu}}.
\end{align}
In the present problem we thus introduced two replication levels. Each replica is hence characterized by a set of two indices: the first $a$ index runs from $1$ to $s$ and specifies the disorder replica, the second $\alpha$ index, running from $1$ to $\beta$ is related to the replication in $\beta$. In total there are therefore $s\times \beta$ replicas. The teacher is set as replica $0$. Implicitly henceforth $a\alpha$ when summed over will be running over $[1,s]\times[1,\beta]\cup {0}$. But
\begin{align}\mathbb{E}_{\bm{F}}\prod\limits_{a\alpha}\prod\limits_{\mu}P_{\mathrm{out}}(y^{\mu}|\bm{F}^{\mu}\cdot\bm{x}^{a\alpha})^{\sigma^{a}_{\mu}}=&\int \prod\limits_{\mu}\prod\limits_{a\alpha}dh^{\mu}_{a\alpha}(\mathrm{det}2\pi\bm{Q})^{-\frac{\alpha N}{2}}e^{-\frac{1}{2}\sum\limits_{\mu}\sum\limits{a\alpha,c\gamma}h^{\mu}_{a\alpha}(\bm{Q}^{-1})^{a\alpha~c\gamma}h^{\mu}_{c\gamma}}\nonumber\\&\int \prod\limits_{a\alpha\ne c\gamma}dq_{a\alpha,c\gamma}\int \prod\limits_{a\alpha\ne c\gamma}d\hat{q}_{a\alpha,c\gamma}e^{\sum\limits_{a\alpha\ne c\gamma}\hat{q}_{a\alpha,c\gamma}(\bm{x}^{a\alpha}\cdot\bm{x}^{c\gamma}-Nq_{a\alpha,c\gamma}}\nonumber\\&\prod\limits_{a\alpha}\prod\limits_{\mu}P_{\mathrm{out}}(y^{\mu}|h^{\mu}_{a\alpha})^{\sigma^{a}_{\mu}}
,\end{align}
where we defined $h^{\mu}_{a\alpha}\equiv\bm{F}^{\mu}\cdot\bm{x}^{a\alpha}$ Gaussian because of the central limit theorem and enforced the definition of its covariance matrix $\bm{Q}$ with integral representations of Dirac deltas. The conjugate matrix is $\hat{\bm{Q}}$. Matrix elements are noted with small $q$. Then 
\begin{align}
\mathbb{E}&_{\bm{F},\bm{x}^{0}}\Xi^{s}=\int d\hat{\bm{Q}}d\bm{Q}e^{-N\mathrm{Tr}\hat{\bm{Q}}\bm{Q}}\left(\int\prod\limits_{a\alpha}d\bm{x}^{a\alpha}P_{X}(\bm{x}^{a\alpha})e^{\sum\limits_{a\alpha\ne c\gamma}x^{a\alpha}\hat{q}_{a\alpha,c\gamma}x^{c\gamma}}\right)^{N}\nonumber\\&\left(\sum\limits_{S^{a}}e^{\phi(\sum\limits_{a}S^{a}~-1)}\int dy \prod\limits_{a\alpha}dh_{a\alpha}(\mathrm{det}2\pi\bm{Q})^{-\frac{1}{2}}e^{-\frac{1}{2}\sum\limits_{a\alpha,c\gamma}h_{a\alpha}(\bm{Q}^{-1})^{a\alpha~c\gamma}h_{c\gamma}}\prod\limits_{a\alpha}P_{\mathrm{out}}(y^{\mu}|h_{a\alpha})^{S^{a}}\right)^{\alpha N},
\end{align}
where we factorized both in $i$ indices (first parenthesis) and in $\mu$ indices (second parenthesis). The free entropy defined in (\ref{eq:GLMmeasure}) then reads
\begin{equation}
\label{eq:GLMPhi}
\Phi(\beta,\phi)=\underset{s\rightarrow 0}{\mathrm{lim}}\frac{1}{s}\underset{\hat{\bm{Q}},\bm{Q}}{\mathrm{extr}}\left\{-\mathrm{Tr}\hat{\bm{Q}}\bm{Q}+\mathrm{ln}I_{X}(\bm{\hat{Q}})+\alpha\mathrm{ln}I_{Y}(\bm{Q})\right\},
\end{equation}
with
\begin{align}
\label{eq:IX}
    &I_{X}=\int\prod\limits_{a\alpha}d\bm{x}^{a\alpha}P_{X}(\bm{x}^{a\alpha})e^{\sum\limits_{a\alpha\ne c\gamma}x^{a\alpha}\hat{q}_{a\alpha,c\gamma}x^{c\gamma}},\\
\label{eq:IY}
    &I_{Y}=\sum\limits_{S^{a}}e^{\phi(\sum\limits_{a}S^{a}~-1)}\int dy \prod\limits_{a\alpha}dh_{a\alpha}(\mathrm{det}2\pi\bm{Q})^{-\frac{1}{2}}e^{-\frac{1}{2}\sum\limits_{a\alpha,c\gamma}h_{a\alpha}(\bm{Q}^{-1})^{a\alpha~c\gamma}h_{c\gamma}}\prod\limits_{a\alpha}P_{\mathrm{out}}(y^{\mu}|h_{a\alpha})^{S^{a}}.
\end{align}

\subsection{Replica Symmetric (RS) ansatz}
The extremization in equation (\ref{eq:GLMPhi}) is hard to carry out. As is now standard in the disordered systems literature we can reduce the number of parameters to be extremized over by enforcing the so-called Replica Symmetric (RS) ansatz (\cite{MezardVirasoro}) on both replication levels
\begin{align}
\label{eq:ansatz}
&q^{0,0}=r^{0},~~\hat{q}^{0,0}=\hat{r}^{0}\\
&q^{a\alpha,0}=m,~~\hat{q}^{a\alpha,0}=\hat{m}\\
\label{eq:r}
&q^{a\alpha,a\alpha}=r,~~\hat{q}^{a\alpha,a\alpha}=-\frac{1}{2}\hat{r}\\
&q^{a\alpha,a\gamma}=Q,~~\hat{q}^{a\alpha,a\gamma}=\hat{Q}\\
\label{eq:ansatz2}
&q^{a\alpha,c\gamma}=q,~~\hat{q}^{a\alpha,c\gamma}=\hat{q},\end{align}
where $q<Q$. Physically, the ansatz (\ref{eq:ansatz})-(\ref{eq:ansatz2}) means that two replicas seeing the same realisation of disorder (i.e., possessing the \textit{same} first index) have an overlap $Q$ greater than the overlap between students seeing different realisations (an thus possessing \textit{different} $a$-index). The $-\frac{1}{2}$ in the definition of $\hat{r}$ (\ref{eq:r}) is just introduced for latter convenience. 

Note finally that while the ansatz (\ref{eq:ansatz}) to (\ref{eq:ansatz2}) is replica-symmetric for both replications, it gives a set of equations that are formally those of a 1RSB problem (\cite{RSB}). This is also a reason why taking 1RSB ansatz (\cite{MezardVirasoro}) in the present large deviation calculation would be rather involved as it would lead to equations in the usual 2RSB form that are numerically involved to be solved.

We plug the RS ansatz (\ref{eq:ansatz})-(\ref{eq:ansatz2}) into the three contributions that make up equation (\ref{eq:GLMPhi}). The trace term is 
\begin{equation}
\label{eq:trace}
 -\mathrm{Tr}\hat{\bm{Q}}\bm{Q}=-\hat{r}^{0}r^{0}-\beta s m\hat{m}+\frac{1}{2}\beta s r\hat{r}-s\frac{\beta(\beta-1)}{2}Q\hat{Q}-\beta^{2}\frac{s(s-1)}{2}q\hat{q}\end{equation}

We can decompose the exponent in (\ref{eq:IX}) according to the ansatz (\ref{eq:ansatz})-(\ref{eq:ansatz2})
\begin{align}
\sum\limits_{a\alpha\ne c\gamma}x^{a\alpha}\hat{q}_{a\alpha,c\gamma}x^{c\gamma}=&\hat{r}^{0}(x^{0})^{2}+\hat{m}x^{0}\sum\limits_{a\alpha \ne 0}x^{a\alpha}-\frac{\hat{r}+\hat{Q}}{2}\sum\limits_{a\alpha\ne0}(x^{a\alpha})^{2}+\frac{\hat{Q}-\hat{q}}{2}\sum\limits_{a\ne 0}\sum\limits_{\alpha,\gamma}x^{a\alpha}x^{a\gamma}\nonumber\\&+\frac{\hat{q}}{2}\sum\limits_{a\alpha\ne0,c\gamma\ne 0}x^{a\alpha}x^{c\gamma}.
\end{align}
In the last but one term index $0$ does not intervene. Introducing Hubbard-Stratonovitch fields $\{\lambda_{a}\}$ for the last but one term and Hubbard-Stratonovith field $\xi$ for the last, $I_{X}$ reads 
\begin{equation}
\label{eq:IXpass}
I_{X}=\int D\xi \int dx^{0}\overline{P_{X}}(x^{0})e^{\hat{r}^{0}(x^{0})^{2}}\left[\int D\lambda\left(\int dx P_{X}(x)e^{\hat{m}x^{0}x-\frac{\hat{r}+\hat{Q}}{2}x^{2}+\sqrt{\hat{Q}-\hat{q}}\lambda x+\sqrt{\hat{q}}\xi x}\right)^{\beta}\right]^{s}.\end{equation}


To carry out the computation for $I_{Y}$ (equation (\ref{eq:IY})) we need to explicitly compute the inverse of the Parisi matrix $\bm{Q}$ involved in equation (\ref{eq:IY}). This is done in the following subsection.

\paragraph*{Some linear algebra for hierarchical matrices}
Name $\bm{\tilde{Q}}\equiv\bm{Q}^{-1}$ the inverse of the overlap matrix $\bm{Q}$. Since $\tilde{\bm{Q}}$ is clearly of the same form as $\bm{Q}$, we can parametrize its coefficient in an identical fashion as those of $\bm{Q}$ $\tilde{r}^{0},\tilde{m},\tilde{r},\tilde{q},\tilde{Q}$.  $\bm{\tilde{Q}}\bm{Q}=\mathbbm{1}_{\beta s+1}$ means 
\begin{align}
&r^{0}\tilde{r}^{0}+\beta s m\tilde{m}=1\\
&r^{0}\tilde{m}+m\tilde{r}+(\beta-1)\tilde{Q}m+\beta(s-1)\tilde{q}m=0\\
&\tilde{r}^{0}m+\tilde{m}(r+(\beta-1)Q+\beta(s-1)q)=0\\
&m\tilde{m}+r\tilde{r}+(\beta-1)Q\tilde{Q}+\beta(s-1)q\tilde{q}=1\\
&m\tilde{m}+r\tilde{Q}+\tilde{r}Q+(\beta-2)Q\tilde{Q}+\beta(s-1)q\tilde{q}=0\\
&m\tilde{m}+r\tilde{q}+(\beta-1)Q\tilde{q}+q\tilde{r}+(\beta-1)q\tilde{Q}+\beta(s-2)q\tilde{q}=0,
\end{align}
yielding
\begin{equation}
\label{eq:inv1}
\tilde{r}^{0}=\frac{r+(\beta-1)Q+\beta(s-1)q}{r^{0}(r+(\beta-1)Q+\beta(s-1)q)-\beta s m^{2}}\end{equation}
\begin{equation}
\tilde{m}=\frac{-m}{r^{0}(r+(\beta-1)Q+\beta(s-1)q)-\beta s m^{2}}\end{equation}
\begin{align}
\tilde{r}&=\frac{\beta m^{2}(q-2Q=r)+\beta((1-s)q^{2}+Q(3Q-2r)+(s-2)q(2Q-r))r^{0}}{(Q-r)(r+(\beta-1)Q-\beta q)(r^{0}(r+(\beta-1)Q+\beta(s-1)q)-\beta s m^{2})}\nonumber\\&+\frac{\beta^{2}(q-Q)(-m^{2}s+((-1+s)q+Q)r^{0})+(Q-r)(m^{2}+(-2Q+r)r^{0})}{(Q-r)(r+(\beta-1)Q-\beta q)(r^{0}(r+(\beta-1)Q+\beta(s-1)q)-\beta s m^{2})}
\end{align}
\begin{equation}
\tilde{Q}=\frac{(Q-r)(m^{2}-Qr^{0})+\beta(q-Q)(m^{2}s-((s-1)q+Q)r^{0}}{(Q-r)(r+(\beta-1)Q-\beta q)(r^{0}(r+(\beta-1)Q+\beta(s-1)q)-\beta s m^{2})}\end{equation}
\begin{equation}
\label{eq:inv2}
\tilde{q}=\frac{-m^{2}+qr^{0}}{(r+(\beta-1)Q-\beta q)(r^{0}(r+(\beta-1)Q+\beta(s-1)q)-\beta s m^{2})}.
\end{equation}
We'll also need the determinant of $\bm{Q}$. To do this, the simplest way is to guess the eigenvectors. For $(x,1,1,...1)^{T}$ we get two eigenvalues $\lambda_{\pm}$ whose product is
\begin{equation}
\lambda_{+}\lambda_{-}=r^{0}(r+(\beta-1)Q+(s-1)\beta q)-\beta s m^{2}.\end{equation}
Then  come $s(\beta-1)$ eigenvectors $\bm{e_{i}}-\bm{e_{i+1}}, ~i\not\equiv 0[\beta]$ (we are indexing starting from $0$), with associated eigenvalues $(r-Q)$. Then for $0\leq s\leq s-2$, $\sum\limits_{k=s\beta+1}^{(s+1)\beta}\bm{e_{k}}-\sum\limits_{k=(s+1)\beta+1}^{(s+2)\beta}\bm{e_{k}}$ is an eigenvector with eigenvalue $r+(\beta-1)Q-\beta q$. Then
\begin{equation}
\mathrm{ln~det}\bm{Q}=\mathrm{ln}(r^{0}(r+(\beta-1)Q+(s-1)\beta q)-\beta m^{2})+(s-1)\mathrm{ln}(r+(\beta-1)Q-\beta q)+(\beta-1)s\mathrm{ln}(r-Q).\end{equation}
The same equality holds with tilde quantities in the right-hand side provided the signs are inverted, since $\mathrm{ln~det}\bm{Q}=-\mathrm{ln~det}\tilde{\bm{Q}}$. Identifying term by term results straightforwardly in a set of relations between tilde and non-tilde quantities (henceforth referred as determinant relations),
\begin{align}
\label{eq:det1}
    &r^{0}(r+(\beta-1)Q+(s-1)\beta q)-\beta m^{2}=[\tilde{r}^{0}(\tilde{r}+(\beta-1)\tilde{Q}+(s-1)\beta \tilde{q})-\beta \tilde{m}^{2}]^{-1}\\
    &r+(\beta-1)Q-\beta q=[\tilde{r}+(\beta-1)\tilde{Q}-\beta \tilde{q}]^{-1}\\
\label{eq:det2}
    &r-Q=[\tilde{r}-\tilde{Q}]^{-1}.
\end{align}
Now decomposing the exponent in $I_{Y}$ (\ref{eq:IY})
\begin{align}
\label{eq:pass3}
-\frac{1}{2}\sum\limits_{a\alpha,c\gamma}h_{a\alpha}\tilde{q}_{a\alpha,c\gamma}h_{c\gamma}=&-\frac{1}{2}\tilde{r}^{0}(h^{0})^{2}-\tilde{m}h^{0}\sum\limits_{a\alpha\ne0}h_{a\alpha}-\frac{\tilde{r}-\tilde{Q}}{2}\sum\limits_{a\alpha\ne 0}(h_{a\alpha})^{2}\nonumber\\&+(\tilde{q}-\tilde{Q})\sum\limits_{a}\sum\limits_{\alpha,\gamma}h_{a\alpha}h_{a\gamma}-\tilde{q}\sum\limits_{a\alpha\ne0,c\gamma\ne0}h_{a\alpha}h_{c\gamma}.
\end{align}
Introducing HS fields $\{\zeta_{a}\}$ and $\eta$ for  the last two sums of (\ref{eq:pass3}) and factorizing in the index $a$ 
\begin{align}
\label{eq:IYpass}
I_{Y}=&\frac{1}{\sqrt{\mathrm{det}2\pi\bm{Q}}}\int dy\int D\eta \int dh^{0}\overline{P_{\mathrm{out}}}(y|h^{0})e^{-\frac{1}{2}\tilde{r}^{0}(h^{0})^{2}}\nonumber\\&\left[\sum\limits_{S=0,1}\int D\zeta e^{\phi S}\left(\int dh P_{\mathrm{out}}(y|h)^{S}e^{-\tilde{m}h^{0}h-\frac{\tilde{r}-\tilde{Q}}{2}h^{2}+\sqrt{\tilde{q}-\tilde{q}}\zeta h+\sqrt{-\tilde{q}}\eta h}\right)^{\beta}\right]^{s}.\end{align}
Now that all terms are computed the next step is then to divide by $s$ and take the $s\rightarrow 0$ limit as prescribed by the replica trick (\ref{eq:replica}). First, we need to enforce that all non-vanishing order $0$ contribution cancel out, since the free entropy should not be diverging. Then, one needs to actually compute first order terms that will contribute in $\Phi$ (\ref{eq:GLMPhi}).

\paragraph*{Order $0$}At order $0$, $I_{Y}=1$ since 
\begin{align}
\underset{s\rightarrow 0}{\mathrm{lim}}\mathrm{ln~det}(2\pi\bm{Q})&=\frac{1}{2}\mathrm{ln}(2\pi \tilde{r}^{0})\nonumber\\&=-\mathrm{ln} \int dy\int D\eta \int dh^{0}\overline{P_{\mathrm{out}}}(y|h^{0})e^{-\frac{1}{2}\tilde{r}^{0}(h^{0})^{2}}.\end{align}
The cancellation of order $0$ terms imposes
\begin{equation}
0=\hat{r}^{0}r^{0}+\mathrm{ln}\int dx^{0}\overline{P_{X}}(x^{0})e^{\hat{r}^{0}(x^{0})^{2}},\end{equation}
where the first term comes from the trace term (\ref{eq:trace}). It follows that $\hat{r}^{0}=0$. Moreover,  because of the saddle point equality
\begin{equation}
q_{a\alpha,c\gamma}=\frac{\int \prod\limits_{d\delta}(dx^{d\delta}P_{X}(x^{d\delta}))x^{a\alpha}x^{c\gamma}e^{\sum\limits_{d\delta\ne e \epsilon}x^{d\delta}\hat{q}_{d\delta,e\epsilon}x^{e\epsilon}}}{\int \prod\limits_{d\delta}(dx^{d\delta}P_{X}(x^{d\delta}))e^{\sum\limits_{d\delta\ne e \epsilon}x^{d\delta}\hat{q}_{d\delta,e\epsilon}x^{e\epsilon}}},\end{equation}
derived straightforwardly from (\ref{eq:GLMPhi}) we also have
\begin{equation}
r^{0}=\int dx^{0}\overline{P_{X}}(x^{0})(x^{0})^{2}.\end{equation}

\paragraph*{Order $1$} 
\paragraph{$I_{X}$}  Carrying out a change of variables $\xi\rightarrow \xi+\hat{q}^{-\frac{1}{2}}\hat{m}x^{0}$ in equation (\ref{eq:IXpass}), $I_{X}$ assumes the compact form
\begin{equation}
\underset{s\rightarrow 0}{\mathrm{lim}}\frac{1}{s}I_{X}=\int D\xi I_{X}^{0}(\xi)\mathrm{ln}I_{X}^{1}(\xi),\end{equation}
where
\begin{align}
&I_{X}^{0}(\xi)=\int dx^{0}\overline{P_{X}}(x^{0})e^{-\frac{\hat{m}^{2}}{2\hat{q}}(x^{0})^{2}+\frac{\hat{m}}{\sqrt{\hat{q}}}\xi x^{0}}\\
&I_{X}^{1}(\xi)=\int D\lambda \left[\int dx P_{X}(x)e^{-\frac{\hat{r}+\hat{Q}}{2}x^{2}+(\sqrt{\hat{Q}-\hat{q}}\lambda+\sqrt{\hat{q}}\xi)x}\right]^{\beta}.\end{align}

\paragraph{$\bm{I_{Y}}$} Changing $\eta\rightarrow  \eta-\frac{\tilde{m}}{\sqrt{-\tilde{q}}}h^{0}$ in (\ref{eq:IYpass}) yields 
\begin{equation}
I_{Y}=\frac{1}{\sqrt{\mathrm{det}2\pi\bm{Q}}}\int dy\int D\eta g^{0}(y,\eta)(g^{1}(y,\eta))^{s},
\end{equation}
with
\begin{align}
&g^{0}(y,\eta)= \int dh^{0}\overline{P_{\mathrm{out}}}(y|h^{0})e^{-\frac{1}{2}(\tilde{r}^{0}-\frac{\tilde{m}^{2}}{\tilde{q}})(h^{0})^{2}-\frac{\tilde{m}}{\sqrt{-\tilde{q}}}h^{0}\eta}\\
&g^{1}(y,\eta)=\sum\limits_{S=0,1}\int D\zeta e^{\phi S}\left(\int dh P_{\mathrm{out}}(y|h)^{S}e^{-\frac{\tilde{r}-\tilde{Q}}{2}h^{2}+\sqrt{\tilde{q}-\tilde{Q}}\zeta h+\sqrt{-\tilde{q}}\eta h}\right)^{\beta}.
\end{align}
Expanding $\mathrm{ln}I_{Y}$ to $\mathcal{O}(s)$ (subscripts in parentheses standing for order in $s$) gives
\begin{align}
\frac{1}{s}\mathrm{ln}I_{Y}=-\frac{1}{2}\mathrm{ln}(\mathrm{det}2\pi\bm{Q})_{(1)}+\frac{1}{\sqrt{2\pi r^{0}}}\int D\eta\int dy\left[g^{0}_{(1)}(y,\eta)+g^{0}_{(0)}(y,\eta)\mathrm{ln}g^{1}_{(0)}(y,\eta)\right].
\end{align}
We used the fact that at order $0$ terms canceled, and the identity $\int D\eta \int dy g^{0}(y,\eta)=\sqrt{2\pi r^{0}}+\mathcal{O}(s)$. But 
\begin{align}
\frac{1}{\sqrt{2\pi r^{0}}}\int D\eta\int dy g^{0}_{(1)}(y,\eta)&=\underset{s\rightarrow 0}{\mathrm{lim}}\frac{1}{\sqrt{2\pi r^{0}}}\partial_{s}\left(\int D\eta\int dy g^{0}(y,\eta)\right)\nonumber\\&= \underset{s\rightarrow 0}{\mathrm{lim}}\frac{1}{\sqrt{2\pi r^{0}}}\partial_{s}\sqrt{\frac{2\pi}{\tilde{r}^{0}}}\nonumber\\&=\frac{-\beta m^{2}}{2r^{0}(r+(\beta-1)Q-\beta q)},
\end{align}
thus
\begin{align}
\label{eq:pass1}
\frac{1}{s}\mathrm{ln}I_{Y}=&-\frac{1}{2}\frac{\beta q}{r+(\beta-1)Q-\beta q}-\frac{1}{2}\mathrm{ln}(r+(\beta-1)Q-\beta q)-\frac{1}{2}(\beta-1)\mathrm{ln}(r-Q)\nonumber\\&+\frac{1}{\sqrt{2\pi r^{0}}}\int dy\int D\eta g^{0}_{(0)}(y,\eta)\mathrm{ln}g^{1}_{(0)}(y,\eta).\end{align}
It's actually possible to proceed to Gaussian changes of variables in the last term so as to exactly cancel the first three contributions in (\ref{eq:pass1}). To do this 
\begin{align}
&h\rightarrow \sqrt{r-Q}h+(r-Q)(\sqrt{\tilde{q}-\tilde{Q}}\zeta+\sqrt{-\tilde{q}}\eta)\\
&\zeta \rightarrow \frac{1}{\sqrt{1-\beta(r-Q)(\tilde{q}-\tilde{Q}}}\zeta -\frac{\beta\sqrt{-\tilde{q}(\tilde{q}-\tilde{Q})}(r-Q)}{1-\beta(r-Q)(\tilde{q}-\tilde{Q}}\eta,
\end{align}
(we used the determinant relations (\ref{eq:det1})-(\ref{eq:det2})) allowing to rewrite the last term in (\ref{eq:pass1}) as
\begin{align}
\label{eq:pass2}
\frac{1}{\sqrt{2\pi r^{0}}}\int dy&\int D\eta  g^{0}_{(0)}(y,\eta)\mathrm{ln}g^{1}_{(0)}(y,\eta)=\frac{\beta}{2}\mathrm{ln}(r-Q)-\frac{1}{2}\mathrm{ln}(1-\beta(r-Q)(\tilde{q}-\tilde{Q}))\nonumber \\&+\int \frac{dy}{\sqrt{2\pi r^{0}}}\int D\eta g^{0}_{(0)}(y,\eta)\mathrm{ln}\left(1+e^{\phi }\int D\zeta \left(\int dh P_{\mathrm{out}}(y|*)\right)^{\beta}\right)\nonumber\\&+\frac{1}{2}\left(-\beta(r-Q)\tilde{q}+\frac{\beta^{2}(-\tilde{q}(\tilde{q}-\tilde{Q})(r-Q)^{2}}{1-\beta(r-Q)(\tilde{q}-\tilde{Q}}\right)\int D\eta \eta^{2}\int  \frac{dy}{\sqrt{2\pi r^{0}}}g^{0}_{(0)}(y,\eta),\end{align}
with 
\begin{equation}
*=\sqrt{r-Q}h+(r-Q)\left(\sqrt{\frac{\tilde{q}-\tilde{Q}}{1-\beta(r-Q)(\tilde{q}-\tilde{Q})}}\zeta +\frac{\sqrt{-\tilde{q}}}{1-\beta(r-Q)(\tilde{q}-\tilde{Q})}\eta\right).\end{equation}
It is straightforward to see
\begin{equation}
\int D\eta \eta^{2}\int  \frac{dy}{\sqrt{2\pi r^{0}}}g^{0}_{(0)}(y,\eta)=\frac{\tilde{r}^{0}\tilde{q}-\tilde{m}^{2}}{\tilde{r}^{0}\tilde{q}},\end{equation}
so the last term in (\ref{eq:pass2}) is 
\begin{align}
\frac{1}{2}\left(-\beta(r-Q)\tilde{q}+\frac{\beta^{2}(-\tilde{q}(\tilde{q}-\tilde{Q})(r-Q)^{2}}{1-\beta(r-Q)(\tilde{q}-\tilde{Q}}\right)\int D\eta \eta^{2}&\int  \frac{dy}{\sqrt{2\pi r^{0}}}g^{0}_{(0)}(y,\eta)\nonumber\\&=-\frac{\beta}{2}\frac{\tilde{r}^{0}\tilde{q}-\tilde{m}^{2}}{\tilde{r}^{0}(\tilde{r}+(\beta-1)\tilde{Q}-\beta\tilde{q})}\nonumber\\&=\frac{1}{2}\frac{\beta q}{r+(\beta-1)Q-\beta q}.
\end{align}
Similarly the first terms in (\ref{eq:pass2}) are 
\begin{align}
\frac{\beta}{2}\mathrm{ln}(r-Q)-\frac{1}{2}\mathrm{ln}(1-\beta(r-Q)(\tilde{q}-\tilde{Q}))&=-\frac{\beta-1}{2}\mathrm{ln}(\tilde{r}-\tilde{Q})-\frac{1}{2}\mathrm{ln}(\tilde{r}+(\beta-1)\tilde{Q}-\beta\tilde{q})\nonumber\\&= \frac{1}{2}\mathrm{ln}(r+(\beta-1)Q-\beta q)+\frac{1}{2}(\beta-1)\mathrm{ln}(r-Q).\end{align}
We again used the determinant identity (\ref{eq:det1})-(\ref{eq:det2}) in the last line. Then tilde quantities in the $s=0$ limit can be accordingly be replaced by their expressions (\ref{eq:inv1})-(\ref{eq:inv2})
\begin{align}
&\underset{s\rightarrow 0}{\mathrm{lim}}\tilde{r}^{0}-\frac{\tilde{m}^{2}}{\tilde{q}}=\frac{q}{qr^{0}-m^{2}}\\
&\underset{s\rightarrow 0}{\mathrm{lim}}\frac{\tilde{m}}{\sqrt{-\tilde{q}}}=\sqrt{\frac{m^{2}}{r^{0}(qr^{0}-m^{2})}}\\
&\underset{s\rightarrow 0}{\mathrm{lim}}\sqrt{\frac{\tilde{q}-\tilde{Q}}{1-\beta(r-Q)(\tilde{q}-\tilde{Q})}}=\frac{\sqrt{Q-q}}{r-Q}\\
&\underset{s\rightarrow 0}{\mathrm{lim}}\frac{\sqrt{-\tilde{q}}}{1-\beta(r-Q)(\tilde{q}-\tilde{Q})}=\sqrt{\frac{qr^{0}-m^{2}}{r^{0}(r-Q)^{2}}}
\end{align}
Ultimately some changes of variables can be used to bring $g^{0}_{(0)}$ to a more compact form
\begin{align}
&h\rightarrow \sqrt{\frac{qr^{0}-m^{2}}{q}}h^{0}+\sqrt{\frac{m^{2}(qr^{0}-m^{2})}{qr^{0}}},~~\eta \rightarrow \sqrt{\frac{qr^{0}}{qr^{0}-m^{2}}}\eta.
\end{align}
Finally
\begin{equation}
\underset{s\rightarrow 0}{\mathrm{lim}}\frac{1}{s}I_{Y}=\int D\eta \int dy I^{0}_{Y}(y,\eta) \mathrm{ln}I^{1}_{Y}(y,\eta),\end{equation}
with
\begin{align}
&I^{0}_{Y}(y,\eta)=\int Dh^{0}\overline{P_{\mathrm{out}}}(y|\sqrt{\frac{qr^{0}-m^{2}}{q}}h^{0}+\sqrt{\frac{m^{2}}{q}}\eta),\\
&I^{1}_{Y}(y,\eta)=1+e^{\phi}\int D\zeta \left(\int dh P_{\mathrm{out}}(y|\sqrt{r-Q}h+\sqrt{Q-q}\zeta +\sqrt{q}\eta))\right)^{\beta}.
\end{align}

\paragraph*{Replica symmetric free entropy for GLM}
Putting everything together the replica free entropy (\ref{eq:GLMPhi}) reads
\begin{align}
\label{eq:finalGLMPhi}
\Phi_{\rm RS} =\underset{\hat{m},\hat{r},\hat{q},\hat{Q},r,q,Q}{\mathrm{extr}}\Big\{&-\beta m\hat{m}+\frac{\beta}{2}r\hat{r}-\frac{\beta(\beta-1)}{2}Q\hat{Q}+\frac{\beta^{2}}{2}q\hat{q}+\int D\xi I_{X}^{0}(\xi)\mathrm{ln}I_{X}^{1}(\xi)\nonumber\\&+\alpha \int D\eta \int dy I^{0}_{Y}(y,\eta) \mathrm{ln}I^{1}_{Y}(y,\eta)\Big\}\end{align}
\begin{align}
&r^{0}=\int dx^{0}\overline{P_{X}}(x^{0})(x^{0})^{2}\\
&I_{X}^{0}(\xi)=\int dx^{0}\overline{P_{X}}(x^{0})e^{-\frac{\hat{m}^{2}}{2\hat{q}}(x^{0})^{2}+\frac{\hat{m}}{\sqrt{\hat{q}}}\xi x^{0}}\\
&I_{X}^{1}(\xi)=\int D\lambda \left[\int dx P_{X}(x)e^{-\frac{\hat{r}+\hat{Q}}{2}x^{2}+(\sqrt{\hat{Q}-\hat{q}}\lambda+\sqrt{\hat{q}}\xi)x}\right]^{\beta}\\
&I^{0}_{Y}(y,\eta)=\int Dh^{0}\overline{P_{\mathrm{out}}}(y|\sqrt{\frac{qr^{0}-m^{2}}{q}}h^{0}+\sqrt{\frac{m^{2}}{q}}\eta)\\
&I^{1}_{Y}(y,\eta)=1+e^{\phi}\int D\zeta \left(\int dh P_{\mathrm{out}}(y|\sqrt{r-Q}h+\sqrt{Q-q}\zeta +\sqrt{q}\eta))\right)^{\beta}
\end{align}

\section{Specialization to perceptron}

\label{appendix:per}

The Bayes-optimal teacher-student setting for the perceptron is defined by the following measures (see section \ref{def} in the main text) 
\begin{align}
\label{eq:priorper}
&\overline{P_{X}}(x)=P_{X}(x)=\frac{1}{\sqrt{2\pi}}e^{\frac{-1}{2}x^{2}}\\
\label{eq:posteriorper}
&\overline{P_{\mathrm{out}}}(y|h)=P_{\mathrm{out}}(y|h)=\delta(y-\mathrm{sgn}(h))
\end{align}

\subsection{Replica symmetric free entropy for the perceptron}
We shall simply plug into the generic GLM (\ref{eq:finalGLMPhi}) expressions the particular priors and posteriors for the perceptron (\ref{eq:priorper})-(\ref{eq:posteriorper}). First,
\begin{align}
I_{X}^{0}(\xi)&=\int \frac{1}{\sqrt{2\pi}}dx^{0}e^{-\frac{1}{2}(1+\frac{\hat{m}^{2}}{\hat{q}})(x^{0})^{2}+\frac{\hat{m}}{\sqrt{\hat{q}}}\xi x^{0}}\\
&=\frac{1}{\sqrt{2\pi(1+\frac{\hat{m}^{2}}{\hat{q}})}}e^{\frac{1}{2}\frac{\hat{m}^{2}}{\hat{q}+\hat{m}^{2}}\xi^{2}},
\end{align}
while two straightforward Gaussian integrals yield
\begin{align}
I_{X}^{1}(\xi)&=\int D\lambda \left[\int \frac{1}{\sqrt{2\pi}} dx P_{X}(x)e^{-\frac{1}{2}(\hat{r}+\hat{Q})x^{2}+(\sqrt{\hat{Q}-\hat{q}}\lambda+\sqrt{\hat{q}}\xi)x}\right]^{\beta}\\
&=\int d\lambda \frac{1}{(1+\hat{r}+\hat{Q})^{\frac{\beta}{2}}}e^{-\frac{1}{2}(1-\beta\frac{\hat{Q}-\hat{q}}{1+\hat{r}+\hat{Q}})\lambda^{2}+\beta\frac{\sqrt{\hat{q}(\hat{Q}-\hat{q})}}{1+\hat{r}+\hat{Q}}\xi\lambda+\frac{\beta}{2}\frac{\hat{q}}{1+\hat{r}+\hat{Q}}\xi^{2}}\\
&=(1+\hat{r}+\hat{Q})^{\frac{1-\beta}{2}}(1+\hat{r}-(\beta-1)\hat{Q}+\beta\hat{q})^{-\frac{1}{2}}e^{\frac{\beta \hat{q}}{2(1+\hat{r}-(\beta-1)\hat{Q}+\beta\hat{q})}\xi^{2}}.
\end{align}
Thus
\begin{equation}
\int D\xi I_{X}^{0}(\xi)\mathrm{ln}I_{X}^{1}(\xi)=-\frac{\beta-1}{2}\mathrm{ln}(1+\hat{r}+\hat{Q})-\frac{1}{2}\mathrm{ln}(1+\hat{r}-(\beta-1)\hat{Q}+\beta\hat{q})+\frac{\beta}{2}\frac{ \hat{q}+\hat{m}^{2}}{1+\hat{r}-(\beta-1)\hat{Q}+\beta\hat{q}}
\end{equation}
Now defining the special function $H(x)\equiv\frac{1}{\sqrt{2\pi}}\int\limits_{x}^{\infty}Dt$

\begin{equation}
I^{0}_{Y}(y,\eta)=\int Dh^{0}\delta(y-\mathrm{sgn}(\sqrt{\frac{q-m^{2}}{q}}h^{0}+\sqrt{\frac{m^{2}}{q}}\eta))=H\left(-y\sqrt{\frac{m^{2}}{q-m^{2}}}\eta\right).
\end{equation}
In writing so we took into account the fact that that $y=\pm1$, which implies also to replace the integral over $y$ in the energetic part by a sum over $\{\pm1\}$. Furthermore
\begin{equation}
I^{1}_{Y}(y,\eta)=1+e^{\phi}\int D\zeta H\left(-\frac{y}{\sqrt{r-Q}}(\sqrt{Q-q}\zeta+\sqrt{q}\eta)\right)^{\beta},
\end{equation}
from which it follows that the energetic term in equation (\ref{eq:finalGLMPhi}) reads
\begin{align}
\alpha \int D\eta \sum\limits_{y=\pm1}I^{0}_{Y}(y,\eta) \mathrm{ln}I^{1}_{Y}(y,\eta)=&2\alpha\int D\eta H\left(-\sqrt{\frac{m^{2}}{q-m^{2}}}\eta\right)\nonumber\\&\mathrm{ln}\left[1+e^{\phi}\int D\zeta H\left(-\frac{1}{\sqrt{r-Q}}(\sqrt{Q-q}\zeta+\sqrt{q}\eta)\right)^{\beta}\right].
\end{align}
The $y=1$ and $y=-1$ being equal modulo a double change of variable $\zeta,\eta\rightarrow -\zeta,-\eta$ whence the factor $2$. Thus for the perceptron
\begin{align}
\label{eq:Phipercept}
\Phi_{\rm RS} =&\underset{\hat{m},\hat{r},\hat{q},\hat{Q},r,q,Q}{\mathrm{extr}}\bigg\{\frac{\beta}{2}r\hat{r}-\beta m\hat{m}-\frac{\beta(\beta-1)}{2}Q\hat{Q}+\frac{\beta^{2}}{2}q\hat{q}-\frac{\beta-1}{2}\mathrm{ln}(1+\hat{r}+\hat{Q})\nonumber\\
&-\frac{1}{2}\mathrm{ln}(1+\hat{r}-(\beta-1)\hat{Q}+\beta\hat{q})+\frac{\beta}{2}\frac{ \hat{q}+\hat{m}^{2}}{1+\hat{r}-(\beta-1)\hat{Q}+\beta\hat{q}}\nonumber\\
&+2\alpha\int D\eta H\left(-\sqrt{\frac{m^{2}}{q-m^{2}}}\eta\right)\mathrm{ln}\left[1+e^{\phi}\int D\zeta H\left(-\frac{1}{\sqrt{r-Q}}(\sqrt{Q-q}\zeta+\sqrt{q}\eta)\right)^{\beta}\right]\bigg\}
\end{align}

\subsection{Saddle-point equations for the perceptron}
The canonical way of carrying out the extremization in equation (\ref{eq:Phipercept}) is to take the saddle-point equations (zero-gradient conditions) and to solve them. The saddle point equations associated to $\Phi(\beta, \phi)$ (equation (\ref{eq:Phipercept})) read
\begin{align}
\label{eq:SP1}
&m^{t}=\frac{\hat{m}^{t}}{1+\hat{r}^{t}-(\beta-1)\hat{Q}^{t}+\beta\hat{q}^{t}}\\
&q^{t}=\frac{\hat{q}^{t}+(\hat{m}^{t})^{2}}{(1+\hat{r}^{t}-(\beta-1)\hat{Q}^{t}+\beta\hat{q}^{t})^{2}}\\
&Q^{t}=\frac{\hat{q}^{t}+(\hat{m}^{t})^{2}}{(1+\hat{r}^{t}-(\beta-1)\hat{Q}^{t}+\beta\hat{q}^{t})^{2}}+\frac{1}{\beta}\frac{1}{1+\hat{r}^{t}-(\beta-1)\hat{Q}^{t}+\beta\hat{q}^{t}}-\frac{1}{\beta}\frac{1}{1+\hat{r}^{t}+\hat{Q}^{t}}\\
&r^{t}=\frac{\hat{q}^{t}+(\hat{m}^{t})^{2}}{(1+\hat{r}^{t}-(\beta-1)\hat{Q}^{t}+\beta\hat{q}^{t})^{2}}+\frac{1}{\beta}\frac{1}{1+\hat{r}^{t}-(\beta-1)\hat{Q}^{t}+\beta\hat{q}^{t}}+\frac{\beta-1}{\beta}\frac{1}{1+\hat{r}^{t}+Q^{t}}\\
&X^{t}=-\frac{1}{\sqrt{r^{t}-Q^{t}}}(\sqrt{Q^{t}-q^{t}}\zeta+\sqrt{q^{t}}\eta)\\
&\hat{q}^{t+1}=2\alpha \int D\eta H\left(-\sqrt{\frac{(m^{t})^{2}}{q^{t}-(m^{t})^{2}}}\eta\right)\frac{e^{2\phi}}{2\pi(r^{t}-Q^{t})}\left[\frac{\int D\zeta H(X^{t})^{\beta-1}e^{-\frac{1}{2}(X^{t})^{2}}}{1+e^{\phi}\int D\zeta H(X^{t})^{\beta}}\right]^{2}\\
&\hat{m}^{t+1}=2\alpha\int d\eta\frac{1}{(2\pi)^{\frac{3}{2}}\sqrt{r-Q}}\frac{1}{\sqrt{1-\frac{(m^{t})^{2}}{q^{t}}}}e^{-\frac{1}{2}\frac{\eta^{2}}{1-\frac{(m^{t})^{2}}{q^{t}}}}\frac{e^{\phi}\int D\zeta H(X^{t})^{\beta-1}e^{-\frac{1}{2}(X^{t})^{2}}}{1+e^{\phi}\int D\zeta H(X^{t})^{\beta}}\\
&\hat{Q}^{t+1}=\frac{2\alpha}{r^{t}-Q^{t}} \int D\eta H\left(-\sqrt{\frac{(m^{t})^{2}}{q^{t}-(m^{t})^{2}}}\eta\right)\frac{\int D\zeta H(X^{t})^{\beta-2}e^{-(X^{t})^{2}}\frac{1}{2\pi}}{1+e^{\phi}\int D\zeta H(X^{t})^{\beta}}\\
\label{eq:SP2}
&\hat{r}^{t+1}=-\frac{2\alpha}{r^{t}-Q^{t}} \int D\eta H\left(-\sqrt{\frac{(m^{t})^{2}}{q^{t}-(m^{t})^{2}}}\eta\right)\frac{\int D\zeta H(X^{t})^{\beta-1}X^{t}e^{-\frac{1}{2}(X^{t})^{2}}\frac{1}{\sqrt{2\pi}}}{1+e^{\phi}\int D\zeta H(X^{t})^{\beta}}.
\end{align}
In practice, equations (\ref{eq:SP1})-(\ref{eq:SP2}) are iterated until convergence. The time indices are derived from an independent computation using Approximate Message Passing (\cite{review}, \cite{ASP}), not shown here. They indicate in which order the equations ought to be iterated in order to converge. Remark that the update schedule very simply consists in updating in parallel all order parameters $m,q,Q,r$ then all auxiliary (hatted) order parameters $\hat{m},\hat{q},\hat{Q},r$. After convergence the order parameters can be used in equation (\ref{eq:Phipercept_main}) to evaluate the free entropy $\Phi$ and subsequently evaluate the complexity $\Sigma(n,v)$ by inverting the Legendre transform (\ref{eq:Legendre}).

\begin{figure}
\begin{center}
\includegraphics[width=0.75\textwidth]{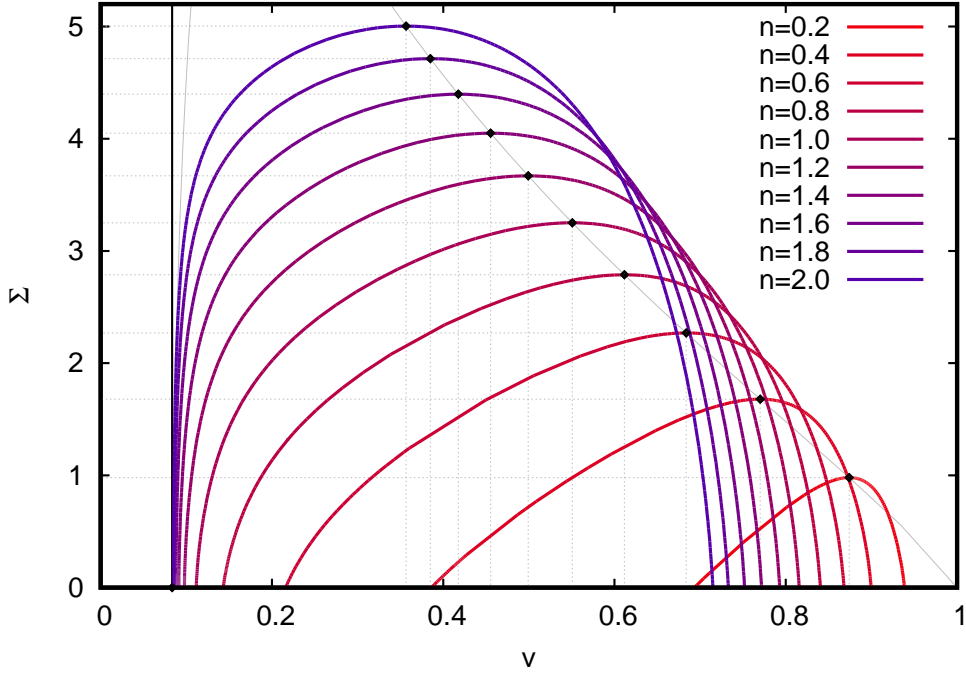} 
\caption{\label{fig:energy-complexity_alpha10}
Complexity $\Sigma(n,v)$ as a function of the Gardner volume $v$ for budgets $n\le 2$ at pool size $\alpha=10$ extracted from the large deviation computations, see also Fig.\,\ref{fig:energy-complexity} for the same curves at different pool size. For any budget $n$ the maximum complexity corresponds to the volume reached by random subset selection (passive learning), see discussion in section \ref{LD} in the main text. We invite the reader to notice that the qualitative allure of the complexity curves remains significantly unchanged as the pool size $\alpha$ is varied (see Fig.\,\ref{fig:energy-complexity} for $\alpha=3$.)}
\end{center}
\end{figure}

We present for illustration in Fig.\,\ref{fig:energy-complexity_alpha10} the complexity curves for $\alpha=10$ for some budgets $n$, and refer the interested reader to Fig.\,\ref{fig:energy-complexity} for the same plot at pool size $\alpha=3$.  As the budget $n$ is increased, smaller values of Gardner volumes become accessible, provided sufficiently informative subset are found. A detailed discussion can further be found in section \ref{LD} in the main text.

\section{Optimal generalization error for the large deviation perceptron}
\label{appendix:gen}
We derive here the expression for the optimal generalization error $\epsilon_{g}$ (in the Bayesian sense) associated to a subset of volume $v$ and of budget $n$ as a function of the perceptron order parameters $m$ and $q$, see appendix \ref{appendix:replica} and
\ref{appendix:per}. The Bayesian $\epsilon_{g}$ was introduced for example in \cite{Baldassi2015SubdominantDC} and, unlike the test error yielded by training the perceptron on some loss, is independent of the training procedure and thus may serve as a nice measure of informativeness for subsets. For the usual perceptron model, it is known that the optimal generalization error is achieved when the student classification is performed by averaging the predicted label over the student measure $P_{X}(\cdot)P_{\mathrm{out}}({\bm{Y}|\bm{F}\cdot})$ and taking the sign thereof. Note that the average predicted label is but the output magnetization $\bm{m}_{\mathrm{out}}$ discussed in section \ref{algo} of the main text. Transposition to the large deviation setting, which allows to fix the budget $n$ and the volume $v$, is straightforward provided one averages over the large deviation measure (\ref{eq:GLMmeasure}). 
By definition the test error is the probability that a new sample $\bm{F}_{\mathrm{new}}\overset{d}{=}\mathcal{N}(0,1)$ is correctly classified by the student according to the output magnetization $\mathbb{E}^{\beta,\phi}_{\bm{x}}\mathrm{sgn}(\bm{x}\cdot\bm{F}_{\mathrm{new}})$
\begin{equation}
    1-\epsilon_{g}=\mathbb{E}_{\bm{x}^{0},\bm{Y},\bm{F},\bm{F_{\mathrm{new}}}}\Theta[\mathrm{sgn}(\bm{x}^{0}\cdot\bm{F}_{\mathrm{new}})\mathbb{E}^{\beta,\phi}_{\bm{x}}\mathrm{sgn}(\bm{x}\cdot\bm{F}_{\mathrm{new}})],
\end{equation}
where $\mathbb{E}^{\beta,\phi}_{\bm{x}}$ denote the average with respect to the large deviation posterior measure (\ref{eq:GLMmeasure}) with control parameters $\beta$ and $\phi$. Introducing an integral representation for a Dirac delta and expanding the resulting exponential 
\begin{align}
\label{eq:introsums}
    1-\epsilon_{g}&=\mathbb{E}_{\bm{x}^{0},\bm{Y},\bm{F_{\mathrm{new}}}}\int dv d\hat{v}e^{iv\hat{v}}\Theta[\mathrm{sgn}(\bm{x}^{0}\cdot\bm{F}_{\mathrm{new}})v]\sum\limits_{j=0}^{\infty}\frac{(i\hat{v})^{j}}{j!}\mathbb{E}_{\bm{F}}(\mathbb{E}_{\bm{x}}^{\beta,\phi}\mathrm{sgn}(\bm{x}\cdot\bm{F}_{\mathrm{new}}))^{j}\\
    &=\mathbb{E}_{\bm{F_{\mathrm{new}}}}\sum\limits_{j=0}^{\infty}\int dv d\hat{v}e^{iv\hat{v}}\frac{(i\hat{v})^{j}}{j!}\mathbb{E}_{\bm{x}^{0},\bm{Y}}\Theta[\mathrm{sgn}(\bm{x}^{0}\cdot\bm{F}_{\mathrm{new}})v]\mathbb{E}_{\bm{F}}(\mathbb{E}_{\bm{x}}^{\beta,\phi}\mathrm{sgn}(\bm{x}\cdot\bm{F}_{\mathrm{new}}))^{j}.
\end{align}
For any fixed $j$, the computation of $\mathbb{E}_{\bm{x}^{0},\bm{Y}}\Theta[\mathrm{sgn}(\bm{x}^{0}\cdot\bm{F}_{\mathrm{new}})v]\mathbb{E}_{\bm{F}}(\mathbb{E}_{\bm{x}}^{\beta,\phi})^{j}$ is formally very similar to the one detailed in appendix \ref{appendix:replica} and follows the same lines. First notice that using the replica trick $\mathbb{E}_{\bm{x}}^{\beta,\phi}\mathrm{sgn}(\bm{x}\cdot\bm{F}_{\mathrm{new}})$ prescribes to introduce as precedently $\beta s$ replicas 
\begin{align}
    \mathbb{E}_{\bm{x}}^{\beta,\phi}\mathrm{sgn}(\bm{x}\cdot\bm{F}_{\mathrm{new}})&=\underset{s\rightarrow 0}{\mathrm{lim}}\Xi^{s-1}
    \sum\limits_{\{\sigma_{\mu}\}}e^{\phi\sum\limits_{\mu}S_{\mu}}
    \left[\int d\bm{x}P_{X}(\bm{x})\prod\limits_{\mu}
    P_{\mathrm{out}}(y^{\mu}|\bm{F}^{\mu}\bm{x})^{\sigma_{\mu}}\right]^{\beta-1}\nonumber\\&\int d\bm{x}P_{X}(\bm{x})\prod\limits_{\mu}
    P_{\mathrm{out}}(y^{\mu}|\bm{F}^{\mu}\bm{x})^{\sigma_{\mu}}\mathrm{sgn}(\bm{x}\cdot\bm{F}_{\mathrm{new}})\\
    &=\underset{s\rightarrow 0}{\mathrm{lim}}\sum\limits_{\bm{\sigma^{1}},...,\bm{\sigma^{n}}}
    \int \prod\limits_{a=1}^{s}\prod\limits_{\alpha=1}^{\beta}
    d\bm{x}^{a\alpha}P_{X}(\bm{x})
    \prod\limits_{\mu}
    P_{\mathrm{out}}(y^{\mu}|\bm{F}^{\mu}\bm{x}^{a\alpha})^{\sigma^{a}_{\mu}}
    e^{\phi\sum\limits_{a,\mu}\sigma^{a}_{\mu}}\nonumber\\&\mathrm{sgn}(\bm{x^{11}}\cdot\bm{F}_{\mathrm{new}}).
\end{align}
From which it follows that 
\begin{align}
\label{eq:3rep}
    (\mathbb{E}_{\bm{x}}^{\beta,\phi}\mathrm{sgn}(\bm{x}\cdot\bm{F}_{\mathrm{new}}))^{j}&=\underset{s\rightarrow 0}{\mathrm{lim}}\sum\limits_{\underset{1\le l\le j,1\le a\le s}{\bm{\sigma^{la}}}}
    \int \prod\limits_{a=1}^{s}\prod\limits_{l=1}^{j}\prod\limits_{\alpha=1}^{\beta}
    d\bm{x}^{a\alpha}P_{X}(\bm{x}^{la\alpha})
    \prod\limits_{\mu}
    P_{\mathrm{out}}(y^{\mu}|\bm{F}^{\mu}\bm{x}^{la\alpha})^{\sigma^{la}_{\mu}}\nonumber\\&
    e^{\phi\sum\limits_{l,a,\mu}\sigma^{la}_{\mu}}\prod\limits_{l=1}^{j}\mathrm{sgn}(\bm{x^{l11}}\cdot\bm{F}_{\mathrm{new}}).
\end{align}
The net effect is simply to transform the first index $a$ into a double index $(l,a)$, thereby introducing a third level of replication. Pursuing equation (\ref{eq:introsums}),
\begin{align}
    1-\epsilon_{g}&=\underset{s\rightarrow 0}{\mathrm{lim}}\sum\limits_{j=0}^{\infty}\int dv d\hat{v}e^{iv\hat{v}}\frac{(i\hat{v})^{j}}{j!}\mathbb{E}_{\bm{x}^{0},\bm{Y},\bm{F}}\sum\limits_{\underset{1\le l\le j,1\le a\le s}{\bm{\sigma^{la}}}}
    \int \prod\limits_{a=1}^{s}\prod\limits_{l=1}^{j}\prod\limits_{\alpha=1}^{\beta}
    d\bm{x}^{a\alpha}P_{X}(\bm{x}^{la\alpha})\nonumber\\&
    \prod\limits_{\mu}
    P_{\mathrm{out}}(y^{\mu}|\bm{F}^{\mu}\bm{x}^{la\alpha})^{\sigma^{la}_{\mu}}
    e^{\phi\sum\limits_{l,a,\mu}\sigma^{la}_{\mu}}\mathbb{E}_{\bm{F_{\mathrm{new}}}}\prod\limits_{l=1}^{j}\mathrm{sgn}(\bm{x^{l11}}\cdot\bm{F}_{\mathrm{new}})\Theta[\mathrm{sgn}(\bm{x}^{0}\cdot\bm{F}_{\mathrm{new}})v]\\
    \label{eq:pass_gen}
    &=\underset{s\rightarrow 0}{\mathrm{lim}}\sum\limits_{j=0}^{\infty}\int dv d\hat{v}e^{iv\hat{v}}\frac{(i\hat{v})^{j}}{j!}e^{sjN\Phi(\beta,\phi)}\int dh^{0}\prod\limits_{l=1}^{j}dh^{l}\frac{e^{-\frac{1}{2}\bm{h}^{T}\mathcal{Q}^{-1}\bm{h}}}{\sqrt{(2\pi)^{j+1}\mathrm{det}\mathcal{Q}}}\Theta[\mathrm{sgn}(h^{0})v]\prod\limits_{l=1}^{j}\mathrm{sgn}(h^{l}).
\end{align}
In going from the second to the last line we introduced overlap variables $h$ as in appendix \ref{appendix:replica}. The rest of the large deviation measure in equation (\ref{eq:3rep}) factorize into $e^{Nsj\Phi(\beta,\phi)}$, with $\Phi$ the free entropy computed in appendices \ref{appendix:replica} and \ref{appendix:per}, and goes to $1$ as the $s\rightarrow0$ limit is taken. We also introduced an overlap matrix $\mathcal{Q}$ of RS form 
\begin{align}
    &\mathcal{Q}_{00}=r^{0}\\
    &\mathcal{Q}_{0l}=m~~ \forall 1\leq l \le j\\
    &\mathcal{Q}_{ll}=r~~ \forall 1\leq l \le j\\
    &\mathcal{Q}_{lk}=q~~ \forall 1\leq l\ne k \le j.
\end{align}
The order parameters $r^{0},r,m$ and $q$ were defined in the replica ansatz equations (\ref{eq:ansatz})-(\ref{eq:ansatz2}). Note that the relevant overlap is $q$, rather than $Q$, since the $j$-fold replication in equation (\ref{eq:3rep}) affects also the selection variables $\sigma$ and thus the variables $\bm{x}^{l11}$ see different disorders, see appendix \ref{appendix:replica}. The inverse $\tilde{\mathcal{Q}}=\mathcal{Q}^{-1}$ is characterized by the coefficients 
\begin{align}
    &\tilde{r}^{0}=\frac{r+(j-1)q}{r^{0}(r+(j-1)q)-jm^{2}}\\
    &\tilde{r}=\frac{r^{0}(r+(j-2)q)-(j-1)m^{2}}{(r-q)(r^{0}(r+(j-1)q)-jm^{2})}\\
    &\tilde{m}=\frac{-m}{(r^{0}(r+(j-1)q)-jm^{2}}\\
    &\tilde{r}=\frac{m^{2}-r^{0}q}{(r-q)(r^{0}(r+(j-1)q)-jm^{2})}.\\
\end{align}
The last integral in equation (\ref{eq:pass_gen}) can be taken core of in the usual manner, by decomposing the exponent and introducing a Hubbard-Stratonovitch field $\eta$, see for example appendix \ref{appendix:replica}. Similarly the expression can then be factorized in $l$ indices. The result is
\begin{align}
    \int dh^{0}\prod\limits_{l=1}^{j}dh^{l}e^{-\frac{1}{2}\bm{h}^{T}\mathcal{Q}^{-1}\bm{h}}\Theta[\mathrm{sgn}(h^{0})v]&\prod\limits_{l=1}^{j}\mathrm{sgn}(h^{l})=\int \frac{D\eta dh^{0}}{\sqrt{(2\pi)^{j+1}\mathrm{det}\mathcal{Q}}}e^{-\frac{1}{2}\tilde{r}^{0}(h^{0})^{2}}e^{\frac{j}{2}(\sqrt{\frac{-\tilde{q}}{\tilde{r}-\tilde{q}}}\eta-\frac{\tilde{m}}{\sqrt{\tilde{r}-\tilde{q}}}h^{0})^{2}}\nonumber\\
    &\left[1-2H\left(\sqrt{\frac{-\tilde{q}}{\tilde{r}-\tilde{q}}}\eta-\frac{\tilde{m}}{\sqrt{\tilde{r}-\tilde{q}}}h^{0}\right)\right]^{j}\\
    &=\int D\eta Dh^{0}\left[1-2H\left(\sqrt{\frac{m^{2}-qr^{0}}{r^{0}(q-r)}}\eta-\sqrt{\frac{m^{2}}{r^{0}(r-q)}}h^{0}\right)\right]^{j}.
\end{align}
This terminates the computation of $\epsilon_{g}$, since from equation (\ref{eq:pass_gen})
\begin{align}
    1-\epsilon_{g}&=\int D\eta Dh^{0}\Theta[h^{0}(1-2H\left(\sqrt{\frac{m^{2}-qr^{0}}{r^{0}(q-r)}}\eta-\sqrt{\frac{m^{2}}{r^{0}(r-q)}}h^{0}\right))]\\
    &=\int D\eta Dh^{0}\Theta[h^{0}(\sqrt{r^{0}q-m^{2}}\eta-m h^{0})]\\
    &=1-\frac{1}{\pi}\mathrm{cos}^{-1}\frac{m}{\sqrt{q}}.
\end{align}
We used the fact that $x\rightarrow1-2H(x)$ was odd and the fact that $r^{0}=1$ for the perceptron model with Gaussian priors, see appendix \ref{appendix:per}.

\section{Additional numerical confirmation}
\label{appendix:numerics}
We supply numerical evidence for some assumptions made in this work, in particular the replica trick (\ref{eq:replica}) and the use of the Gardner volume as a measure of informativeness (see section \ref{mutualinfo} in the main text). 

\begin{figure}
\begin{center}
\includegraphics[width=0.7\textwidth]{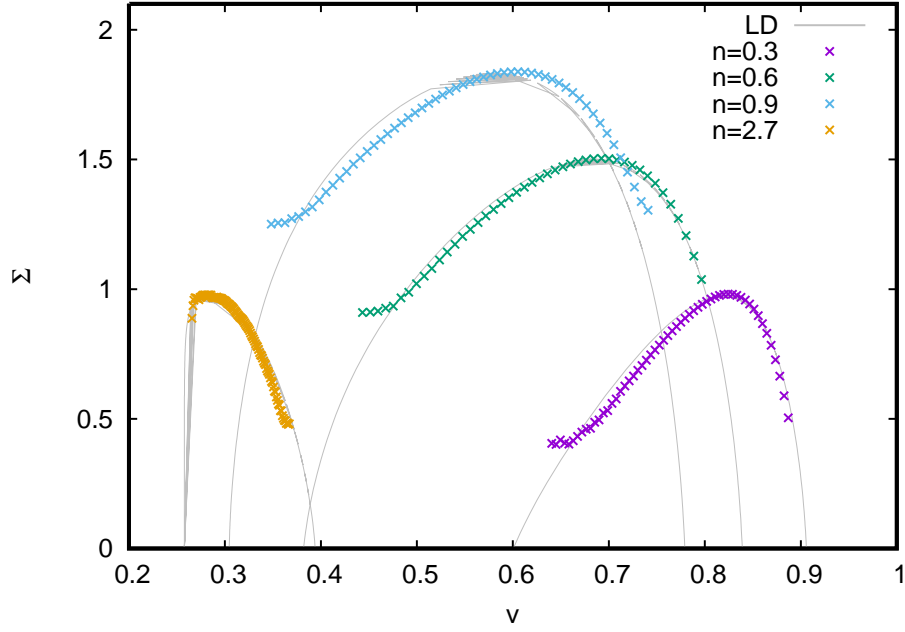}
\caption{\label{fig:scatter} Complexity vs volume curves for $\alpha=3$, and $n\in\{0.3,0.6,0.9,2.7\}$. The dots are the values extracted from numerical experiments performed at $N=20$ by repeatedly sampling passively $10^7$ times a subset of cardinality $n$ out of a fixed pool of size $\alpha=3$. Solid lines are the theoretical complexities as predicted by the large deviation computations, see also Fig.\,\ref{fig:energy-complexity}. Volumes were evaluated using the AMP Algorithm~\ref{alg:TAPeq}. The agreement is rather good knowing the discrepancies that ought to be expected because of running AMP Algorithm~\ref{alg:TAPeq} at finite and small $N$.}
\end{center}
\end{figure}

First, we sample numerically at random subsets of cardinalities $n\in\{0.3,0.6,0.9,2.7\}$ out of a pool of cardinal given by $\alpha=3$, and plot the complexity extracted therefrom in Fig.\,\ref{fig:scatter}. The volumes were evaluated using the AMP algorithm \ref{alg:TAPeq}, and simulations were performed at $N=20$, with $10^{7}$ draws. Because of computational limitations $N$ has been kept small, while AMP is known to be valid only in the $N\uparrow\infty$ limit, hence inducing errors due to finite size. Nevertheless, the agreement with the theoretical curves for $\Sigma(n,v)$ is quite good.

We finally present a numerical check of the theoretical prediction for the $m(v)$ curves, see Fig.\,\ref{fig:energy-magnetization}. At large instance size, $N=2000$, it is not computationally feasible to obtain sufficient statistics for observing the large deviations of the volume through passive subset sampling, as it was done in the previous experiment. Thus, we resorted to the label-informed AL-AMP active learning strategy for biasing the subset selection towards more/less informative subsets. In particular, we constructed each subset by mixing varying ratios of maximally informative samples (selected according to the informed AL-AMP procedure, see algorithm \ref{alg:uncertainty} and Tab.\,\ref{tab:alg_tab}) and minimally informative samples (selected according to the same procedure but with the reversed sorting order). In the figure, the pool size is $\alpha=10$ and the budget is fixed to $n=1.5$. For each subset, the AMP algorithm \ref{alg:TAPeq} was run to get the estimator $\hat{\bm{x}}$ and the magnetization $m$ was deduced therefrom. This incidentally corroborates once more that using the Gardner volume instead of the magnetization to judge for the informativeness of a selection is coherent.  
\begin{figure}
\begin{center}
\includegraphics[width=0.7\textwidth]{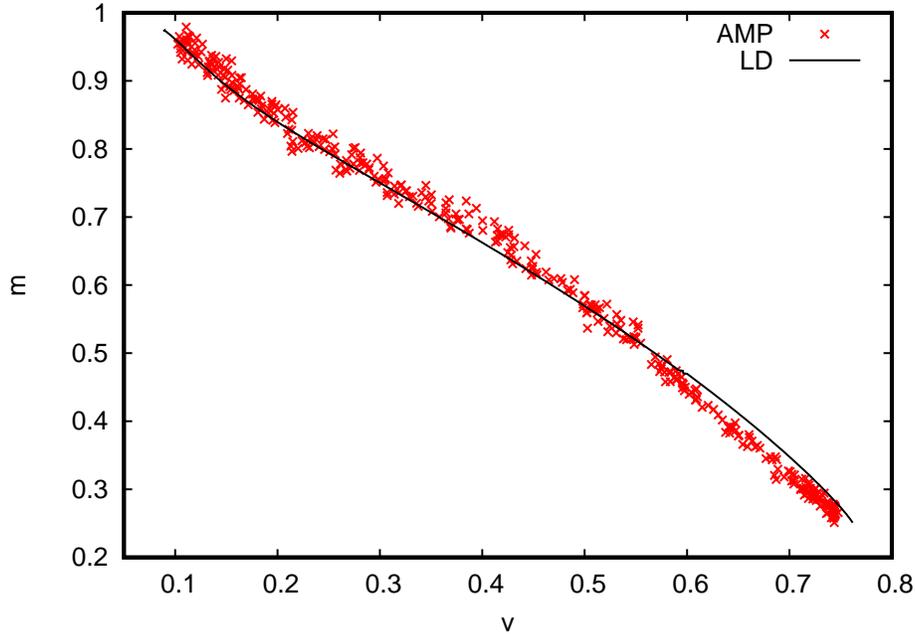} 
\caption{\label{fig:mag_vs_volume}
Magnetization $m$ against Gardner volume $v$ for various subsets. The experiments were performed at system size $N=10^3$, pool size $\alpha=10$ and budget $n=1.5$. Subsets covering a wide range of volumes were designed by varying the ratio of informative samples (using label-informed AL-AMP, see section \ref{algo}) and uninformative samples (selected using simple passive learning). Magnetizations and volumes were evaluated using the AMP procedure \ref{alg:TAPeq}. In solid line is the typical $m(v)$ curve predicted by the large deviation computations, which agrees quite well with the numerical simulations.}
\end{center}
\end{figure}

\end{document}